\documentclass[aps,pre,nofootinbib,amsfonts,superscriptaddress,longbibliography,showkeys,notitlepage]{revtex4-1}
\usepackage[utf8]{inputenc}
\usepackage[T1]{fontenc}
\usepackage[english]{babel}
\usepackage{graphicx,color}

\usepackage{booktabs} 

\usepackage{xcolor}
\usepackage{amsmath}
\usepackage{amssymb}
\usepackage{bm}
\DeclareMathOperator{\aleft}{\mathit{left}}
\DeclareMathOperator{\aright}{\mathit{right}}

\usepackage{subfigure}
\usepackage{color}
\definecolor{Orange}{rgb}{0.9,0.5,0}
\definecolor{NavyBlue}{rgb}{0.1, 0.4, 0.8}
\definecolor{Magenta}{rgb}{0.8, 0.1, 0.6}
\definecolor{Yellow}{rgb}{0.9, 0.8, 0.4}
\definecolor{Red}{rgb}{1, 0,0}

\usepackage[nameinlink,noabbrev]{cleveref}

\usepackage{tabularx}
\usepackage{booktabs}   
\newcommand{\ra}[1]{\renewcommand{\arraystretch}{#1}}

\newcolumntype{R}[1]{>{\raggedleft\let\newline\\\arraybackslash\hspace{0pt}}m{#1}}
\usepackage{threeparttable}
\usepackage{colortbl}

\usepackage[super]{nth}

\DeclareMathOperator{\AppS}{AppS}

\usepackage[ruled]{algorithm2e} 
\usepackage{afterpage}

\begin{document}

\title{Apps, Places and People: strategies, limitations and trade-offs in the physical and digital worlds}

\author{Marco De Nadai}
\affiliation{Vodafone Research, Paddington Central, London, W2 6BY, UK}
\affiliation{{Department of Information Engineering and Computer Science, University of Trento, Via Sommarive, 9I, 38123 Povo (TN), Italy}}
\affiliation{Mobs Lab, Fondazione Bruno Kessler, Via Sommarive 18, 38123 Povo (TN), Italy}

\author{Angelo Cardoso}
\affiliation{Vodafone Research, Paddington Central, London, W2 6BY, UK}

\author{Antonio Lima}
\affiliation{Vodafone Research, Paddington Central, London, W2 6BY, UK}

\author{Bruno Lepri}
\affiliation{Mobs Lab, Fondazione Bruno Kessler, Via Sommarive 18, 38123 Povo (TN), Italy}

\author{Nuria Oliver}
\affiliation{Vodafone Research, Paddington Central, London, W2 6BY, UK}


\begin{abstract}
Cognition has been found to constrain several aspects of human behaviour, such as the number of friends and the number of favourite places a person keeps stable over time. This limitation has been empirically defined in the physical and social spaces. But do people exhibit similar constraints in the digital space? We address this question through the analysis of pseudonymised mobility and mobile application (app) usage data of 400,000 individuals in a European country for six months. Despite the enormous heterogeneity of apps usage, we find that individuals exhibit a conserved capacity that limits the number of applications they regularly use. Moreover, we  find that this capacity steadily decreases with age, as does the capacity in the physical space but with more complex dynamics. 
Even though people might have the same capacity, applications get added and removed over time. In this respect, we identify two profiles of individuals: app \emph{keepers} and \emph{explorers}, which differ in their stable (keepers) vs exploratory (explorers) behaviour regarding their use of mobile applications. 
Finally, we show that the capacity of applications predicts mobility capacity and vice-versa. By contrast, the behaviour of \emph{keepers} and \emph{explorers} may considerably vary across the two domains.
Our empirical findings provide an intriguing picture linking human behaviour in the physical and digital worlds which bridges research studies from Computer Science, Social Physics and Computational Social Sciences.
\end{abstract}

\flushbottom
\maketitle
%
%
\thispagestyle{empty}

\section*{Introduction}
Recent studies on mobility and social interactions suggest that cognitive constraints, rather than time, might be the primary cause of the limited number of places and friends that people maintain at any point in their life time~\cite{miritello2013limited, dunbar2016online, alessandretti2018, alessandretti2018evidence, cinelli2019selective}.
Thanks to the wide adoption of smartphones and the proliferation of mobile applications (apps), almost any human need --from entertainment to social connection or productivity-- can be satisfied by at least one of the two million mobile apps available in the major app stores~\cite{statista}.
As a consequence, people spend an increasing amount of time on their smartphones, reaching an average of 3 hours per day in 2018~\cite{appannie2019} and triggering debates about their effect on human cognition and attention~\cite{hadar2015,loh2015,wilmer2017}. 
Interestingly, despite this ever-growing \emph{digitisation} of human life and availability of apps, people tend to exploit a small set of repeatedly used apps~\cite{falaki2010diversity}. Is it the case then, that human behaviour on digital devices exhibits similar dynamics and constraints as those found in the \emph{physical} world?

Similarly to mobility, we know that human behaviour on mobile phones has regular daily rhythms~\cite{marquez2017not} that coexist with a \emph{bursty} and highly heterogeneous usage~\cite{falaki2010diversity}, where most of the applications struggle to stay relevant longer than a fortnight~\cite{sonntag2013netradar}.
The existing literature has leveraged these findings to predict short-term dynamics (e.g., next used app), understand the relationship with user's actions and context, or recommend apps~\cite{peltonen2018hidden, shin2012understanding, yang2016apps, yu2018smartphone, karatzoglou2012climbing, do2010their}.
Only a few studies have tried to characterise the statistical properties of the adoption and use of mobile  applications~\cite{falaki2010diversity, zhao2016discovering}, relying however on fixed observation windows, which hinder the temporal variations of used and abandoned apps over time. Such a limitation has been mainly caused by the absence of data describing long-term human behaviour on mobile phones. The available data is indeed usually based on a few weeks of network traffic generated by both foreground and background applications, which sometimes are automatically launched by the phone without the user's will \cite{aggarwal2014prometheus}. 

In this paper, we analyse the use of foreground applications to compare human behaviour between the digital and physical worlds. To the best of our knowledge, this is the first research effort to study app usage alongside with mobility in a large population over six months. In a modern society with high adoption of smartphones, understanding applications usage has both theoretical and practical implications in a variety of fields from the design of digital services to human behaviour understanding and modelling.

\section*{Results}
We study six months of pseudonymized data collected through an Android application installed in hundreds of thousands of devices in a European country. Upon installation, the app --which runs in the background-- asks its users for explicit consent to record at regular intervals the state of the device, its usage and the context where it is used (e.g., GPS coordinates). We consider only data generated by users having GPS locations covering at least 80\% of the hours of each individual, and having  application usage data for the entire period. We characterise human mobility through an individual's set of locations, where \emph{locations} are defined as places where people stop for at least 15 minutes. To uniformly analyse human behaviour on mobile devices, we consider only those apps available in the Google Play Store, which is the main store for Android devices.
After the filtering, the data consists of 415,000 users that stop in 138 million locations and use 69,000 different applications that were launched for a total of around 1 billion times. We refer to the Methods and Supplementary Information (SI) for further details about the processing and sampling approaches.

As aforementioned, previous literature has mainly investigated application usage from either limited and controlled contexts, or short-term passive collection of network traffic, which limit the ability to capture app usage.
Network-level measurements, in particular, include data generated by both background and foreground applications, which makes it very hard to analyse actual human behaviour~\cite{aggarwal2014prometheus}. 
Thus, we here begin by describing some statistical properties of the foreground application usage. 

\begin{figure}[t!h]
    \centering
    \includegraphics{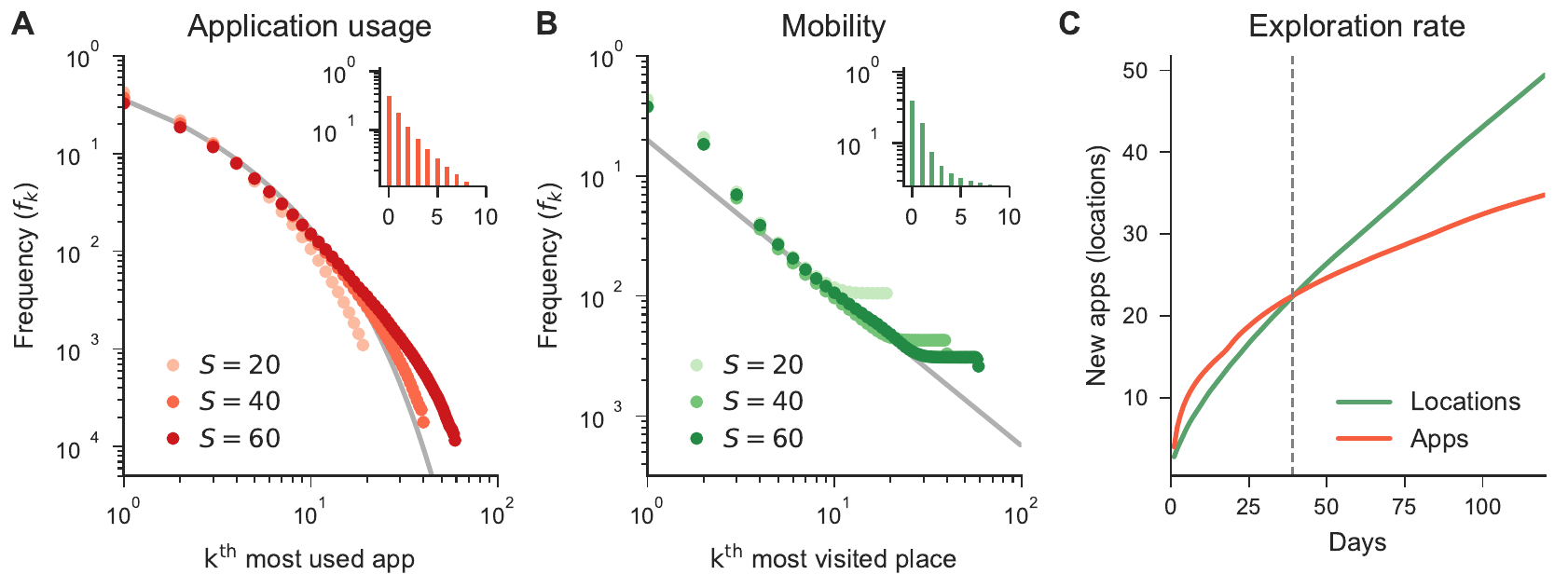}
    \caption{Aggregated statistics of app usage and mobility. (A) Truncated power-law plot showing the frequency $f_k$ of the most used $k$ applications. We show in red three users with a different number of distinct applications $S$. The grey line shows that the empirical data is well approximated by $f_k \sim (k + k_0)^{-\alpha} exp(-k/c) $, where $\alpha = 1.19 \pm 0.01$, $k_0 = 1.14 \pm 0.07$, and $c = 8.32 \pm 0.75$.  The inset illustrates that more than 40\% of the time people are found using the first two apps; (B) Zipf's law plot showing the frequency $f_k$ of the most visited $k$ locations of users with different number of distinct total locations $S$. The grey line shows that the empirical data is well approximated by $f_k \sim (k)^{-\alpha}$, where $\alpha = 1.27 \pm 0.01$.  The inset illustrates that more than 40\% of the time people are found in the first two locations; (C) Average number of new apps and locations discovered by users over time.}
    \label{fig:one}
\end{figure}

During the entire six-month period people used on average a total of 27 different apps, mostly belonging to the \emph{Communication} and \emph{Social} categories (see Supplementary Information (SI) Figure S2 (A)). The \emph{Communication} category includes all the applications that allow users to send messages to other people (e.g., WhatsApp, Messenger), while the \emph{Social} category includes Social Network Apps (e.g., Pinterest, Facebook, Instagram). 
The usage frequency and time spent on apps by an individual is heavily skewed.
We find that the app usage is well described by a truncated power-law, where the frequency $f$ of the $k^{th}$ most visited location is well approximated by: $f_k \sim (k + k_0)^{-\alpha} exp(-k/c) $, with exponent $\alpha = 1.19 \pm 0.01$, $k_0 = 1.14 \pm 0.07$ and a cut off value $c = 8.32 \pm 0.75$. Thus, the time spent by people on phones is mostly focused on a few apps, although users possess at least 26 applications (see SI Figure S7).
\Cref{fig:one} (A) shows the distribution of application usage for people with different number of distinct apps $S$.
Similar results are obtained for background applications, where the distribution is even more skewed towards the first app (see SI Figure S5 (E)).
Mobility is well described by a power law distribution $f_k \sim k^{-\alpha}$ with $\alpha = 1.27 \pm 0.01$, which is compatible to the results found in literature~\cite{song2010modelling} ($\alpha = 1.2 \pm 0.1$). While the distributions between mobility and application usage are different, the exponent $\alpha$ of the power laws show a similar tendency towards the skewed use of time in locations and apps. 

Nevertheless, human behaviour evolves over time. 
\Cref{fig:one} (C) shows the number of new locations and apps that people discover over time. 
We model the total number of apps as $L(t)_{apps} \propto t^{\gamma_1}$ where $t$ is the time and $\gamma_1$ a growing coefficient, and the number of locations as $L(t)_{mob} \propto t^{\gamma_2}$. 
We find $\gamma_2 = 0.64$ and $\gamma_1=0.41$, revealing a surprising fact: people explore new locations over time at a much faster pace than they add new apps (in particular, after 39 days from \Cref{fig:one} (C)). Thus, while people tend to use a small set of applications, they also continuously explore new applications over time at a slower rate than new locations, though.

To explain this apparent contradiction, we characterise the mobility and app usage through the \emph{activity space} and \emph{the app space}.
In mobility, the activity space~\cite{alessandretti2018evidence} is defined as the set $MobS_i(t) = [\mathfrak{l}_1, \mathfrak{l}_2, \ldots, \mathfrak{l}_n]$ of stop locations an individual $i$ visits at least twice and spends on average more than 10 min per week over a time-window $t$. 
In a similar fashion, we define the \emph{app space} as the set of applications $AppS_i (t) = [\mathfrak{a}_1, \mathfrak{a}_2, \ldots, \mathfrak{a}_n]$ that are used at least twice by the user $i$ in a time window $t$. 
The app space describes the set of apps that are used at any point in time by an individual.
As both application usage and mobility are \emph{bursty}~\cite{marquez2017not, song2010limits}, too short and too long time windows might hide dynamics and erroneously identify spurious behaviour. 
Thus, similarly to previous work~\cite{alessandretti2018evidence} we use a time window $d=20$ weeks long. Note that we tested the sensitivity of the window size and found no significant differences.

\subsection*{Capacity and activity of app usage} 
\begin{figure}[ht]
  \centering
  \includegraphics[width=0.95\textwidth]{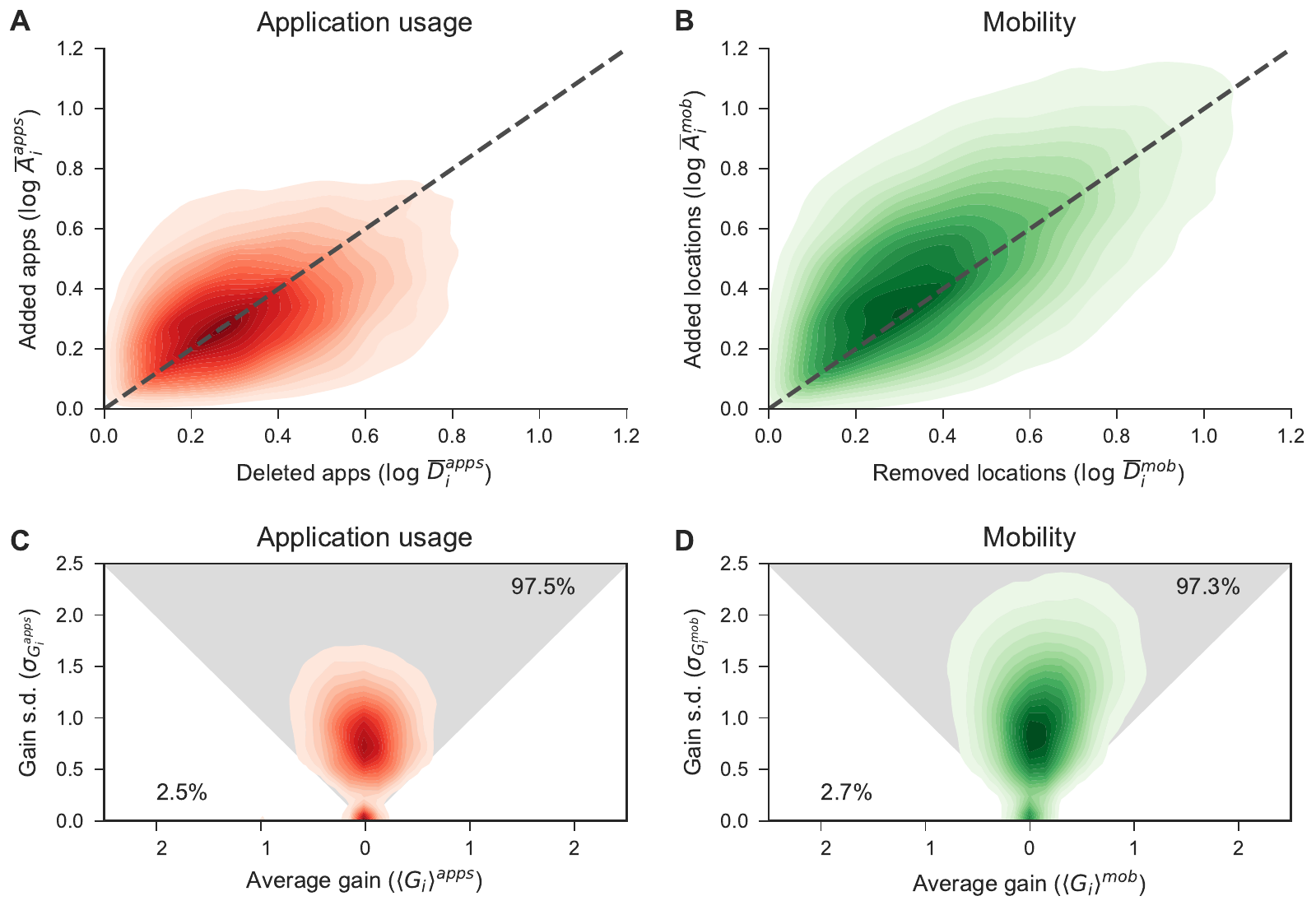}
  \caption{Capacity and activity of application usage and mobility for all the individuals. (A-B) Density plots $\rho (\log{\overline{D}_i}, \log{\overline{A}_i})$ for users with $\log{\overline{A}_i} > 0$ and $\log{\overline{D}_i} > 0$ in both application usage and mobility. (C-D) Average gain of apps (locations) versus the standard deviation of the app (locations) gain. The grey area corresponds to individuals with a conserved size of applications (locations): $\overline{G}_i \leq \sigma_{G_i}$ for user $i$. 97.5\% of the users has a conserved capacity in application usage, and 97.3\% of them has a conserved capacity in mobility.}
  \label{fig:capacity}
\end{figure}

The activity and app spaces allow to observe the evolution of the set of preferred locations and applications over time.
First, for each user $i$ we define the \emph{app capacity} as the number of distinct apps used by the user in a time window $t$: $C_i^{\textit{apps}}(t) = |\AppS_i (t)|$ to observe how the number of familiar apps changes over time.
Then, we model the relative average capacity across the sample population as: $\overline{C}^{\textit{apps}}(t)/\left<\overline{C}\right>^{\textit{apps}} = \alpha^{\textit{apps}} + \beta^{\textit{apps}} t$. 
We find that the app capacity is constant in time, as the slope of this linear relation does not markedly differ from zero ($\beta^{\textit{apps}} = 0.0023$).
We also test the alternative hypothesis where the capacity would be a consequence of the high heterogeneity of the sample population: with some people shrinking it and others expanding it over time. 
For each individual $i$ we measure the \emph{app gain} $G_i^{\textit{apps}}(t) = A_i^{\textit{apps}}(t) - D_i^{\textit{apps}}(t)$, defined as the difference between the number of added $A_i^{\textit{apps}}(t)$ and removed $D_i^{\textit{apps}}(t)$ apps over two time windows (e.g. $[t, t+d)$ and $[t+r, t+r+d)$ with a slide $r$), and we also define the
\emph{net gain} as the average absolute gain over time divided by the standard deviation of it $\left|\left<G_i\right>^{\textit{apps}}\right| / \sigma_{G^{\textit{apps}}_i}$. People having net gain within one standard deviation (s.d.) are expected to be consistent with $\left<G_i\right> = 0$, while people with $\left|\left<G_i\right>^{\textit{apps}}\right| / \sigma_{G^{\textit{apps}}_i} > 1$ increase or decrease their net gain over time.
We find that $97.3\%$ of the people in our data have $\left|\left<G_i\right>^{\textit{apps}}\right| / \sigma_{G_i^{\textit{apps}}} \leq 1$, thus exhibiting a conserved \emph{app capacity} (see \Cref{fig:capacity} (C)).

We also computed  the same metrics for mobility. We find that the mobility capacity $C^{\textit{mob}}$ is constant over time ($\beta^{\textit{mob}} = 0.0064$) and that $97.5\%$ of the users have a net gain that does not significantly differ from zero ($\left|\left<G_i\right>^{\textit{mob}}\right| / \sigma_{G^{\textit{mob}}_i} \leq 1$) (see \Cref{fig:capacity} (D))
These results are in agreement with the literature~\cite{alessandretti2018evidence}. 

Interestingly, this empirical study uncovers a remarkable similarity between mobility and the application usage domains.
\Cref{fig:capacity} (A) depicts the relationship between the average number of added and removed apps, while \Cref{fig:capacity} (B) shows the same relationship for the mobility domain.

By observing the categories of apps that are kept and dropped from the app space the most, we can shed light on the shift in the interests of our users. 
The categories kept for a longer time are \emph{Communication} and \emph{Productivity}, which supports the view of smartphones being used mainly for social connection and productivity. In particular, two apps are on average continuously used and kept in the app space: WhatsApp and Facebook, perhaps due to the network effect, i.e. as the number of people using a service increases so does the value of using it~\cite{katz1994systems, pan2011composite}. Conversely, proprietary and niche apps such as Samsung Keyboard and Secure Folder are dropped from the app space very frequently. 
We refer the reader to SI Section S4 for further details about this.

Our results indicate that the app capacity is conserved for most individuals. However, it might be a direct consequence of time constraints, as common sense would suggest. People have limited time to allocate to different activities on a daily basis. Thus, we break application usage in daily modules and shuffle it through two types of randomisation.  For example, given the temporal sequence of app usage for two different users, the local randomisation shuffles the temporal order of the sequence of apps for one user, while the global randomisation shuffles the sequences across all the users. We refer to the Methods Section for further details.
We find that capacity is constant even after shuffling the individual time series with both types of randomisation. Moreover, the two-sided Kolmogorov-Smirnov (KS)~\cite{ref1} test rejects the hypothesis that the two random time series have a similar underlying distribution to the original one (KS-local: 0.55 $p$-value < 0.001, KS-global: 0.98 $p$-value < 0.001). 
As the KS distance is lower in the case of local randomisation: these results suggest that the app capacity is not just a consequence of time constraints but an inherent property of human behaviour.

Interestingly, we find that human behaviour in the app space is very similar to that of the mobility space, as shown in \Cref{fig:capacity}. 
We find a significant and positive correlation (Spearman's rank $0.21$, $p$-value $< 0.001$) between mobility and app capacity, but also between the individual number of new locations and new apps (Spearman's rank $0.16$, $p$-value $< 0.001$) (see SI Figure S1). 
These positive and significant correlations might be a consequence of a trade-off between mobile phone usage and mobility, where people tend to decrease their mobility when they use the phone for longer times and vice-versa.
Hence, we break individual behaviour into one-day modules, where each module describes the number of locations and the cumulative time spent on apps in the day. Then, we compare human dynamics through the number of visited locations and the total time spent on applications in three different time windows (\emph{i.e.}, daily, weekly, monthly). 
We do not find any negative correlation between these variables, which would imply the existence of a trade-off between mobile phone usage and human mobility. On the contrary, we do find a slightly positive correlation. 
In other words and to our surprise, the higher the mobility, the higher the usage of apps is (see SI Section S2). 

\subsection*{Keepers and Explorers}
Previous work has found that people can be grouped in two groups through the regularity of their behaviour: people who tend to behave according to constant and repetitive habits and those who tend to change their behaviour over time~\cite{riefer2017coherency, pappalardo2015returners, miritello2013limited}.
This result has been found, under different names, in previous work regarding social connections~\cite{miritello2013limited} and  mobility~\cite{pappalardo2015returners}.
However, to the best of our knowledge, the literature has not yet explored this dichotomy in the behaviour regarding the use of applications.

\begin{figure}[ht]
    \centering
    \includegraphics[width=\textwidth]{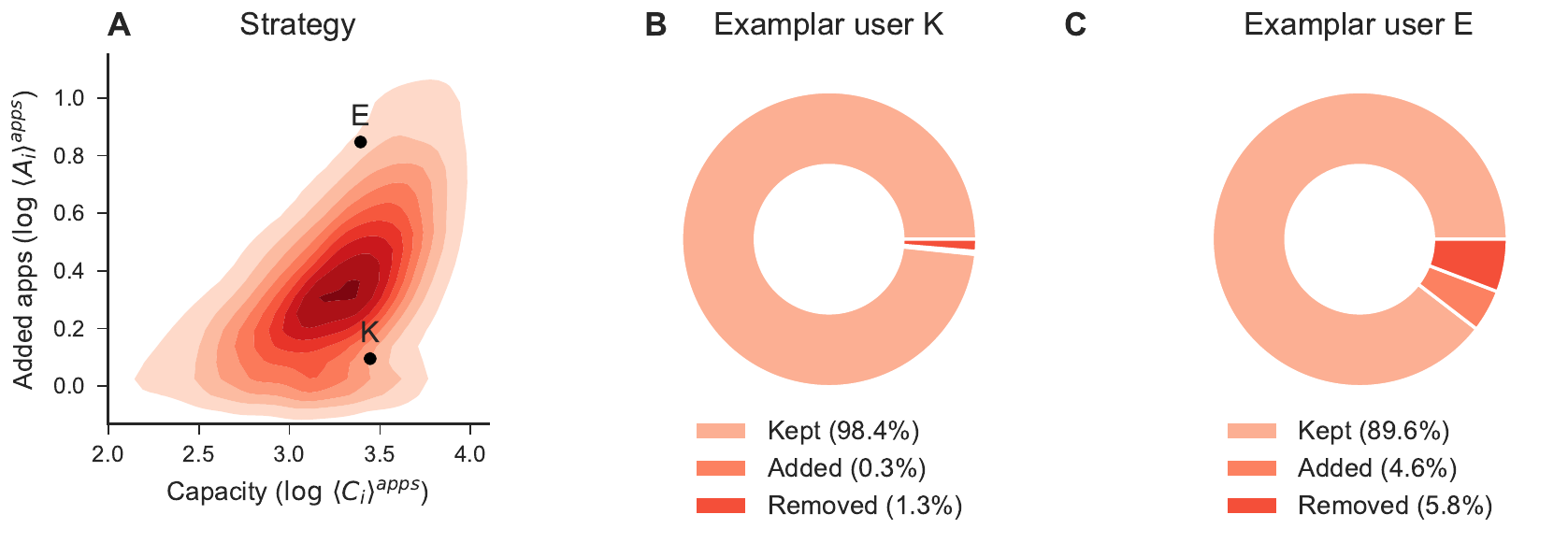}
    \caption{Variability of strategy. (A) Density plot showing between the app capacity and the new apps discovered. We randomly select two users with similar capacity, but different strategy, which we name E and K. (B) Individual K -- an app keeper -- has an average of less than 1\% added and removed apps. (C) Individual E -- an app explorer -- explores more than user K. }
    \label{fig:explorers}
\end{figure}

We note that users with the same app capacity might have a very different rate of new apps discovered. To illustrate this point, we randomly select two users, namely $K$ and $E$, from the set of people who have similar app capacity but exhibit a very different number of newly discovered apps. 
\Cref{fig:explorers} shows that user $K$ used roughly the same applications during the entire period of study, whereas user $E$ added new apps in the app space and removed some of them as well, thus maintaining constant capacity.
Similarly to previous work~\cite{miritello2013limited}, we encode this \emph{strategy} through the ratio between the number of newly adopted apps and the user's average capacity $R^{\textit{apps}}_{i} = \left<A_i\right>^{\textit{apps}} / \left<C_i\right>^{\textit{apps}}$. 
We define application \emph{explorers} to be those users with $R_{i} \gg \beta$ and application \emph{keepers} to be those users with $R_{i} \ll \beta$, where $\beta$ corresponds to the average behaviour over all the users. We compute the same measure in the physical space using the mobility capacity and the new locations added to the users' \emph{activity space}, defining $R_i^{\textit{mob}} = \left<A_i\right>^{\textit{mob}} / \left<C_i\right>^{\textit{mob}}$.
Previous work has defined the explorers-keepers dichotomy in mobility through the radius of gyration, which is the radius of the circumference that encloses most of the locations usually visited by an individual. Thus, such a definition is about the size of the geographic space explored by people. However, our definition of explorers vs keepers is about the rate of adoption of new locations that are visited regularly by individuals. Therefore, our definition is consistent with our concept of explorers vs keepers in the applications domain and also to previous work in the case of social connections~\cite{miritello2013limited}.

By defining explorers from the distribution of $R^{\textit{apps}}$ as those with $R^{\textit{apps}}$ higher than the \nth{80} percentile and returners as those with $R^{\textit{apps}}$ lower than the \nth{20} percentile, we observe that explorers adopt on average one app every 28 weeks ($\overline{A}^{\textit{apps}} = 0.72$), while keepers adopt one new app every 500 weeks ($\overline{A}^{\textit{apps}} = 0.04$).

When we apply the same concept to mobility, the results are surprising. As common sense would suggest, discovering and visiting new locations costs more energy than discovering and installing new mobile applications, even if some of them are not free. However, the number of adopted and discarded locations is larger than the number of apps. On average, mobility explorers embrace a new familiar location every 17 weeks ($\overline{A}^{\textit{mob}} = 1.16$), while keepers adopt a new location every 181 weeks ($\overline{A}^{\textit{mob}} = 0.11$). Although social relations and tightly coupled with mobility~\cite{alessandretti2018, doi:10.1098/rsif.2014.1128}, most of the social relations might be managed by only a few apps such as Facebook, Whatsapp and Messanger, which are the most kept applications in the app space (see SI Table S2).

As both capacity and activity are correlated across domains, we also compare strategies between application usage and mobility.
First, we classify individuals in one of the three classes in the app domain: namely \emph{explorer}, \emph{keeper} and \emph{other}. This last class describes all the "average" behaviour within two standard deviations of $R^{\textit{apps}}$. We do the same according to their mobility strategy. 
Then, we use a Random Forest Classifier with 20 estimators where the independent variable is each user's app strategy $R^{\textit{apps}}$, and we predict the corresponding class in the mobility domain. 
We fit the model in a Stratified 5-fold Cross-Validation fashion to avoid over-fitting.
While the model shows severe imbalance over the average class (\emph{other}), we find that it is possible to predict the mobility strategy using as input the application strategy with an F1-score of $0.54$. 
We obtain similar results when we train using the labels of the users' mobility strategy $R^{\textit{mob}}$ to predict their app strategy (F1-score: $0.53$). Even though the strategies correlate across the app and mobility domains, we find that it is very challenging to predict one from the other.

\subsection*{Age-dependency of app usage and mobility}
In this section, we shift our focus to demographic differences in the users' app and mobility behaviour. Our data contains age information for 92.6\% of the users who range from 18 to 68 years old with $\mu = 39$ years and $\sigma = 12$ years, as shown in the SI Figure S6.

\begin{figure}[ht]
  \centering
  \includegraphics[width=0.95\textwidth]{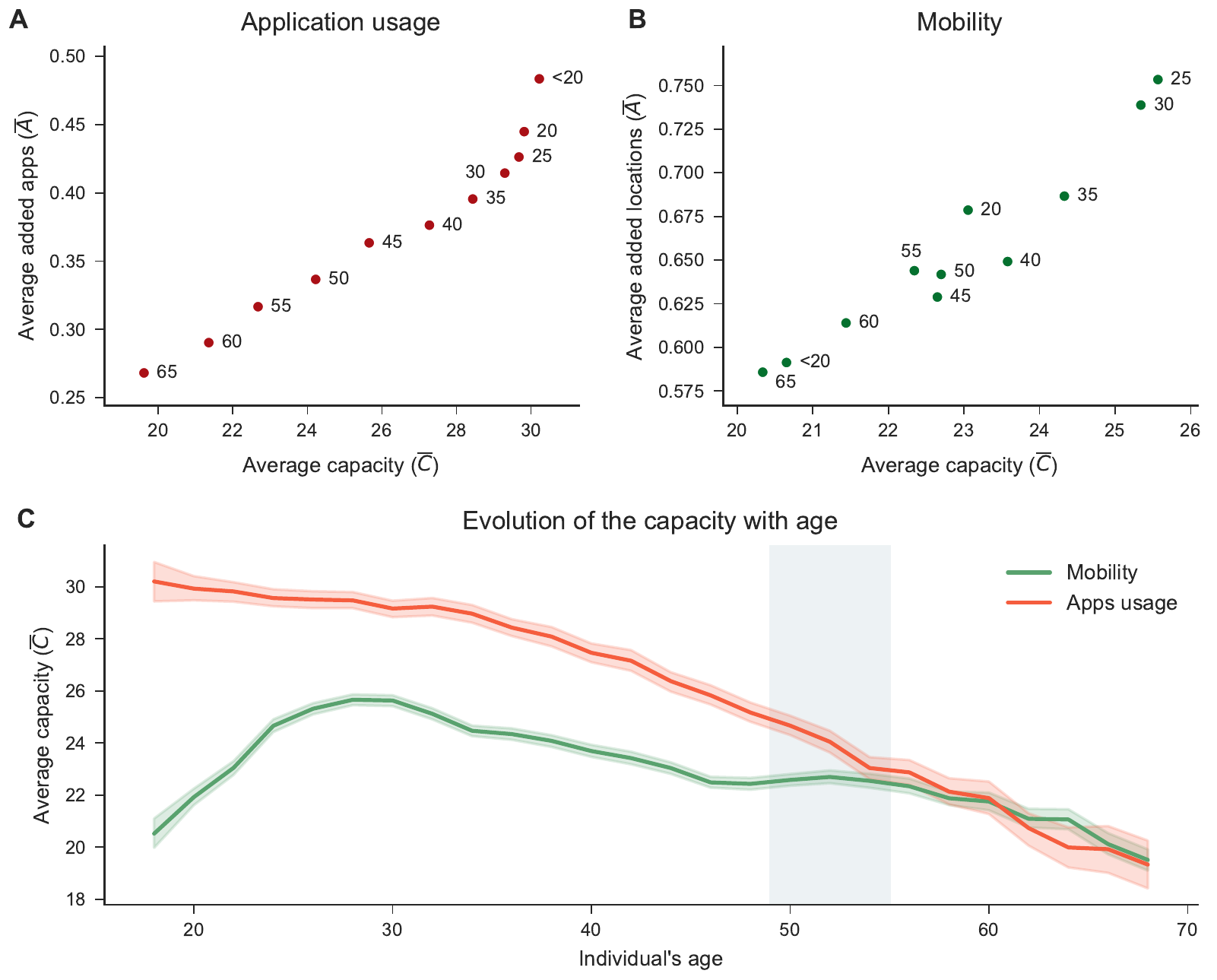}
  \caption{Age-variations of capacity and novelty. (A-B) Average capacity and novelty for groups of users of different age. (A) suggests a strong dependence of applications with age, while (B) is less clear. (C) shows that the average app capacity steadily decreases when age increases, while mobility capacity increases until around 28 years, then decrease until 46 where it remains constant, to decrease again from 56 years onward.}
  \label{fig:individuals}
\end{figure}

We analyse the relationship between the age of the users in our data set, and their app and mobility capacities. As perhaps expected, we find a strong negative correlation between age and app usage. \Cref{fig:individuals} (C) shows that younger people --those aged between 18 and 24 years-- have the highest average number of applications in their app space. From 20 years of age onward, the average app capacity declines monotonically.   
Interestingly, mobility behaves very differently depending on the age of our users. 
As illustrated in \Cref{fig:individuals} (C), in early adulthood individuals seem to increase their mobility capacity from around 20 to 26 preferred locations. Then, a slow decrease starts until approximately the age of 48, where the capacity plateaus for a few years to then decrease again monotonically with age. 
This result might be related to life events that have an impact on people's mobility. A recent survey~\cite{rauch_2018} involving 300,000 Britons, shows that the point in life where people are the most dissatisfied is 49, while the peak of satisfaction is around 30 years. Surprisingly, the results of this survey correspond to the beginning of the plateau in average mobility capacity and the highest point of mobility capacity in our data, respectively.

Additionally, we found that people's strategy is influenced by their age. In the app space, the older the people are, the less prone they are to discover and add new apps in the app space (Spearman-r $-0.99$, $p$-value $< 0.001$). 
In the case of mobility, such a relationship is weaker but still very significant (Spearman-r $-0.61$, $p$-value $< 0.001$). 
\Cref{fig:individuals} (A) shows an evident tendency of older people to have a smaller capacity and exploration rate than younger people, while \Cref{fig:individuals} (B) shows that this relationship is less clear. Thus, we analyse the distribution of strategies in our users grouped by age --in bins of five-years. We find that in mobility the ratio of explorers and returners is almost constant for all groups (see SI Figure S3 (B)). However, in the case of apps we observe that young people have a low percentage of returners and a high rate of explorers ($\sim12\%$ and $\sim28\%$ respectively), and the percentage of returners increases with age (the last age-group has $\sim27\%$ of returners and $\sim19\%$ explorers) (see SI Figure S3 (A))

\section*{Discussion}
In this paper, we have studied whether the inherent properties of human behaviour found in social relations and mobility also apply to mobile application usage. We have compared the statistical properties of app usage and mobility through the analysis of a large data set containing the mobile app behaviour of hundreds of thousands of individuals over six months. 
We have found that, despite the high heterogeneity of application usage, individuals can be described through their \emph{app capacity} and their \emph{app activity space}.
The former expresses the conserved and limited number of apps an individual uses in any point of their life --which remarkably is almost the same value as their mobility capacity. The latter represents the number of novel apps adopted over time.
We have found that app capacity is not a direct consequence of time constraints but an individual behaviour that might be connected with our own cognitive limits, in line with the Dunbar's social brain hypothesis~\cite{Hill2003}, which fixes to around 150 the number of friends people can maintain at any point in their life. In this respect, it is interesting to note that recent preliminary results on online content consumption are aligned with our results~\cite{cinelli2019selective}.

However, people are not all alike. Previous work has found two main strategies grouping people into those who tend to exploit the same items over time, and those who tend to explore new items, where items are relationships, places or actions.
While researchers have referred to these strategies with different terms --\emph{e.g.} returners and explorers~\cite{pappalardo2015returners} and social keepers and explorers~\cite{miritello2013limited}-- the findings are consistent across different domains.
Here, we have found empirical evidence for the first time that people also exhibit this exploration vs exploitation dichotomy in their app usage behaviour.
Moreover, we also provide a novel definition of mobility keepers and explorers, which is consistent in the app and social domains~\cite{miritello2013limited}.
Surprisingly, the strategies do not always match across domains: keepers in the app domain could be explorers in the physical space and \emph{vice-versa}.
Additionally, contrarily to common sense, we also find that the number of adopted and discarded locations is larger than the number of added and discarded apps: mobility explorers adopt on average more than one new location every 17 weeks, whereas app explorers adopt on average one new app every 28 weeks.

Capacity and strategy are correlated with age in both the applications and mobility domains. 
Although we did not analyse the data for the same individuals over multiple decades, our result suggests that capacity is a stable property over short-term periods of time but would evolve (and mainly decrease) with age.
Age-specific life events and goals profoundly influence our behaviour, especially in the social domain~\cite{wrzus2013social}. Also, cognitive abilities --which typically decline with age-- play a role in shaping our interests, actions and our decisions regarding how we spend our time~\cite{holtzman2004social}. 
Although we identify a significant positive correlation between the mobility and the app capacity and also between new locations and new apps, these correlations, are not due to a trade-off between mobile phone usage and mobility. Thus, we do not find clear evidence on the impact of  mobile phone usage on mobility.

A quantitative understanding of human behaviour on digital devices is uttermost important to interpret the profound and fast changes happening in our contemporary society.
Together, our results not only extend to the mobile app domain previous empirical results on social relations~\cite{Saramaki942, miritello2013limited} and human mobility~\cite{alessandretti2018evidence}, but also shed new light on the interplay between the physical and the digital worlds.

\section*{Methods}
We analyse data concerning application usage and GPS coordinates. We use data from ~400,000 users in a European country for six months, ending in July of 2018. To safeguard personal privacy, all data is pseudonymised and collected with full informed consent, in agreement with existing data privacy and data protection regulations and analysed according to our institution's code of conduct. All results and insights are aggregated over thousands of individuals.

From the raw data we obtain two different data streams of the same users to support our analysis: the \emph{User Locations} data stream, composed of (user ID, date, time, latitude, longitude); and the \emph{User Application Usage} data stream, composed of (user ID, date, application name, aggregated time spent, number of times opened). For a subset of the users with also have some limited \emph{User Demographic} data, consisting of their ID and self-reported age for 92.6\% of the users. Note that all user IDs are hashed and randomised to preserve anonymity. 

\mbox{ } \\
\textbf{User Locations.} This data has been obtained from the GPS coordinates that are collected from either actual GPS measurements with an error of less than 30m, or through a WiFi look-up performed by the device's operating system. We do so to avoid spurious detection of locations. 
We filtered out all users whose locations are available less than 80\% of the time, to ensure that we have enough data to characterise their mobility appropriately. 
The resulting data has 383,422 users with mobility information, with a median number of GPS coordinates per day per user of 96.
We extract the \emph{stop events} with an algorithm based on Hariharan and Toyama~\cite{hariharan2004project}, where a stop event is defined as a temporal sequence of GPS coordinates in a radius of $\Delta s$ meters where the user stayed for at least $\Delta t$ minutes. The algorithm, its optimisation and its complexity are explained in details in the SI. 
The presented results are for $\Delta s = 50$ meters and $\Delta t = 15$ minutes, parameters similar to the literature~\cite{alessandretti2018evidence}. 
For each user, we define \emph{stop locations} as the sequences of \emph{stop events} that can be considered part of the same place. 
To determine a \emph{stop location} from \emph{stop events} we use the DB-scan ($minPoints$, $\epsilon$) algorithm~\cite{ester1996density} that groups points within $\epsilon = \Delta s -5$ meters of distance to form a cluster with at least $minPoints = 1$ \emph{event} (see SI Section S1 for more details). 
In sum, we characterise the users' mobility by their sequence of \emph{stop locations}.

\mbox{ } \\
\textbf{Application usage.} This data contains the timestamp, number of launches and the screen time of all the applications that are launched by the users.
This allows the analysis of the real behaviour of the users, without the well-known problems of  network traffic data~\cite{aggarwal2014prometheus}.
To further highlight the differences between background and foreground applications we refer to the SI Figure S5.
We focus our analysis on applications that are downloadable from the Google Play store, thus excluding vendor-specific applications. 
This allows investigating people's behaviour uniformly across devices.
At this moment, the store has 11 main categories, namely \emph{Business}, \emph{Communication}, \emph{Fitness}, \emph{Game}, \emph{Lifestyle}, \emph{Music}, \emph{Personalization}, \emph{Photography}, \emph{Reading}, \emph{Social}, \emph{Tools}, and \emph{Travel}.
We filter out apps belonging to the \emph{Personalization} and \emph{Tools} to avoid most of the manufacturer-specific lock screen apps and custom launchers (e.g. \texttt{com.htc.launcher}).

To reduce the noise of the data, we also exclude those app launches that lasted less than one second, which might be related to apps opened by mistake.
This rule of thumb choice exclude just a small portion of data without affecting the overall results (see the SI).

The resulting data has 92,943 users with app usage information, with a median number of 7 different apps launched per day per user.

\mbox{ } \\
\textbf{Global and local randomisation.}
We test whether the app capacity $C^{\textit{apps}}$ is a consequence of time constraints by applying two randomisation techniques previously proposed for the mobility space~\cite{alessandretti2018evidence}. 
Let $X$ and $Y$ be two users with daily usage of apps $D_X=[d_{X,t}, d_{X,t+1}, \ldots, d_{X,t+n}]$ and $D_Y=[d_{Y,t}, d_{Y,t+1}, \ldots, d_{Y,t+n}]$ the two randomisation strategies are:
\begin{itemize}
    \item \emph{Local}. We permute the order of the sliding observation windows at random. For example we shuffle the two original time-series of $X$ and $Y$: $D_X=[d_{X,t+5}, d_{X,t}, \ldots, d_{X,t+1}]$ and $D_Y=[d_{Y,t}, d_{Y,t+n}, \ldots, d_{Y,t+1}]$.
    \item \emph{Global}. We permute an individual's data across the entire data. For example: $D_X=[d_{Y,t}, d_{X,t+1}, \ldots, d_{Y,t+n}]$ and $D_Y=[d_{Y,t}, d_{Y,t+1}, \ldots, d_{X,t+n}]$ (Note the shuffle appendix of the last element in the sequence).
\end{itemize}

\section*{Acknowledgements}

The authors would like to thank Riccardo di Clemente, Lorenzo Lucchini and Dario Patanè for the insightful discussions and comments.

\section*{Author contributions statement}

M.D.N, A.C., A.L. and N.O. conceived the experiments, M.D.N and A.C. conducted the experiments, M.D.N, A.C., A.L. and N.O. analysed the results.  All authors reviewed the manuscript.

\bibliography{output}

\begin{thebibliography}{39}%
\makeatletter
\providecommand \@ifxundefined [1]{%
 \@ifx{#1\undefined}
}%
\providecommand \@ifnum [1]{%
 \ifnum #1\expandafter \@firstoftwo
 \else \expandafter \@secondoftwo
 \fi
}%
\providecommand \@ifx [1]{%
 \ifx #1\expandafter \@firstoftwo
 \else \expandafter \@secondoftwo
 \fi
}%
\providecommand \natexlab [1]{#1}%
\providecommand \enquote  [1]{``#1''}%
\providecommand \bibnamefont  [1]{#1}%
\providecommand \bibfnamefont [1]{#1}%
\providecommand \citenamefont [1]{#1}%
\providecommand \href@noop [0]{\@secondoftwo}%
\providecommand \href [0]{\begingroup \@sanitize@url \@href}%
\providecommand \@href[1]{\@@startlink{#1}\@@href}%
\providecommand \@@href[1]{\endgroup#1\@@endlink}%
\providecommand \@sanitize@url [0]{\catcode `\\12\catcode `\$12\catcode
  `\&12\catcode `\#12\catcode `\^12\catcode `\_12\catcode `\%12\relax}%
\providecommand \@@startlink[1]{}%
\providecommand \@@endlink[0]{}%
\providecommand \url  [0]{\begingroup\@sanitize@url \@url }%
\providecommand \@url [1]{\endgroup\@href {#1}{\urlprefix }}%
\providecommand \urlprefix  [0]{URL }%
\providecommand \Eprint [0]{\href }%
\providecommand \doibase [0]{http://dx.doi.org/}%
\providecommand \selectlanguage [0]{\@gobble}%
\providecommand \bibinfo  [0]{\@secondoftwo}%
\providecommand \bibfield  [0]{\@secondoftwo}%
\providecommand \translation [1]{[#1]}%
\providecommand \BibitemOpen [0]{}%
\providecommand \bibitemStop [0]{}%
\providecommand \bibitemNoStop [0]{.\EOS\space}%
\providecommand \EOS [0]{\spacefactor3000\relax}%
\providecommand \BibitemShut  [1]{\csname bibitem#1\endcsname}%
\let\auto@bib@innerbib\@empty
\bibitem [{\citenamefont {Miritello}\ \emph {et~al.}(2013)\citenamefont
  {Miritello}, \citenamefont {Lara}, \citenamefont {Cebrian},\ and\
  \citenamefont {Moro}}]{miritello2013limited}%
  \BibitemOpen
  \bibfield  {author} {\bibinfo {author} {\bibfnamefont {Giovanna}\
  \bibnamefont {Miritello}}, \bibinfo {author} {\bibfnamefont {Rub{\'e}n}\
  \bibnamefont {Lara}}, \bibinfo {author} {\bibfnamefont {Manuel}\ \bibnamefont
  {Cebrian}}, \ and\ \bibinfo {author} {\bibfnamefont {Esteban}\ \bibnamefont
  {Moro}},\ }\bibfield  {title} {\enquote {\bibinfo {title} {Limited
  communication capacity unveils strategies for human interaction},}\ }\href
  {\doibase 10.1038/srep01950} {\bibfield  {journal} {\bibinfo  {journal}
  {Scientific reports}\ }\textbf {\bibinfo {volume} {3}},\ \bibinfo {pages}
  {1950} (\bibinfo {year} {2013})}\BibitemShut {NoStop}%
\bibitem [{\citenamefont {Dunbar}(2016)}]{dunbar2016online}%
  \BibitemOpen
  \bibfield  {author} {\bibinfo {author} {\bibfnamefont {R.~I.~M.}\
  \bibnamefont {Dunbar}},\ }\bibfield  {title} {\enquote {\bibinfo {title} {Do
  online social media cut through the constraints that limit the size of
  offline social networks?}}\ }\href {\doibase 10.1098/rsos.150292} {\bibfield
  {journal} {\bibinfo  {journal} {Royal Society Open Science}\ }\textbf
  {\bibinfo {volume} {3}},\ \bibinfo {pages} {150292} (\bibinfo {year}
  {2016})}\BibitemShut {NoStop}%
\bibitem [{\citenamefont {Alessandretti}\ \emph
  {et~al.}(2018{\natexlab{a}})\citenamefont {Alessandretti}, \citenamefont
  {Lehmann},\ and\ \citenamefont {Baronchelli}}]{alessandretti2018}%
  \BibitemOpen
  \bibfield  {author} {\bibinfo {author} {\bibfnamefont {Laura}\ \bibnamefont
  {Alessandretti}}, \bibinfo {author} {\bibfnamefont {Sune}\ \bibnamefont
  {Lehmann}}, \ and\ \bibinfo {author} {\bibfnamefont {Andrea}\ \bibnamefont
  {Baronchelli}},\ }\bibfield  {title} {\enquote {\bibinfo {title}
  {Understanding the interplay between social and spatial behaviour},}\ }\href
  {\doibase 10.1140/epjds/s13688-018-0164-6} {\bibfield  {journal} {\bibinfo
  {journal} {EPJ Data Sci.}\ }\textbf {\bibinfo {volume} {7}},\ \bibinfo
  {pages} {36} (\bibinfo {year} {2018}{\natexlab{a}})}\BibitemShut {NoStop}%
\bibitem [{\citenamefont {Alessandretti}\ \emph
  {et~al.}(2018{\natexlab{b}})\citenamefont {Alessandretti}, \citenamefont
  {Sapiezynski}, \citenamefont {Sekara}, \citenamefont {Lehmann},\ and\
  \citenamefont {Baronchelli}}]{alessandretti2018evidence}%
  \BibitemOpen
  \bibfield  {author} {\bibinfo {author} {\bibfnamefont {Laura}\ \bibnamefont
  {Alessandretti}}, \bibinfo {author} {\bibfnamefont {Piotr}\ \bibnamefont
  {Sapiezynski}}, \bibinfo {author} {\bibfnamefont {Vedran}\ \bibnamefont
  {Sekara}}, \bibinfo {author} {\bibfnamefont {Sune}\ \bibnamefont {Lehmann}},
  \ and\ \bibinfo {author} {\bibfnamefont {Andrea}\ \bibnamefont
  {Baronchelli}},\ }\bibfield  {title} {\enquote {\bibinfo {title} {Evidence
  for a conserved quantity in human mobility},}\ }\href {\doibase
  10.1038/s41562-018-0364-x} {\bibfield  {journal} {\bibinfo  {journal} {Nature
  Human Behaviour}\ }\textbf {\bibinfo {volume} {2}},\ \bibinfo {pages}
  {485--491} (\bibinfo {year} {2018}{\natexlab{b}})}\BibitemShut {NoStop}%
\bibitem [{\citenamefont {Cinelli}\ \emph {et~al.}(2019)\citenamefont
  {Cinelli}, \citenamefont {Brugnoli}, \citenamefont {Schmidt}, \citenamefont
  {Zollo}, \citenamefont {Quattrociocchi},\ and\ \citenamefont
  {Scala}}]{cinelli2019selective}%
  \BibitemOpen
  \bibfield  {author} {\bibinfo {author} {\bibfnamefont {Matteo}\ \bibnamefont
  {Cinelli}}, \bibinfo {author} {\bibfnamefont {Emanuele}\ \bibnamefont
  {Brugnoli}}, \bibinfo {author} {\bibfnamefont {Ana~Lucia}\ \bibnamefont
  {Schmidt}}, \bibinfo {author} {\bibfnamefont {Fabiana}\ \bibnamefont
  {Zollo}}, \bibinfo {author} {\bibfnamefont {Walter}\ \bibnamefont
  {Quattrociocchi}}, \ and\ \bibinfo {author} {\bibfnamefont {Antonio}\
  \bibnamefont {Scala}},\ }\bibfield  {title} {\enquote {\bibinfo {title}
  {Selective exposure shapes the facebook news diet},}\ }\href@noop {}
  {\bibfield  {journal} {\bibinfo  {journal} {arXiv preprint arXiv:1903.00699}\
  } (\bibinfo {year} {2019})}\BibitemShut {NoStop}%
\bibitem [{\citenamefont {{Statista}}(2019)}]{statista}%
  \BibitemOpen
  \bibfield  {author} {\bibinfo {author} {\bibnamefont {{Statista}}},\
  }\href@noop {} {\enquote {\bibinfo {title} {Number of apps available in
  leading app stores as of 3rd quarter 2018},}\ } (\bibinfo {year} {2019}),\
  \bibinfo {note}
  {\url{https://www.statista.com/statistics/276623/number-of-apps-available-in-leading-app-stores/},
  Last accessed on 2019-03-29}\BibitemShut {NoStop}%
\bibitem [{\citenamefont {{App Annie}}(2019)}]{appannie2019}%
  \BibitemOpen
  \bibfield  {author} {\bibinfo {author} {\bibnamefont {{App Annie}}},\
  }\href@noop {} {\emph {\bibinfo {title} {The state of mobile 2019}}},\
  \bibinfo {type} {Tech. Rep.}\ (\bibinfo {year} {2019})\BibitemShut {NoStop}%
\bibitem [{\citenamefont {Hadar}\ \emph {et~al.}(2015)\citenamefont {Hadar},
  \citenamefont {Eliraz}, \citenamefont {Lazarovits}, \citenamefont {Alyagon},\
  and\ \citenamefont {Zangen}}]{hadar2015}%
  \BibitemOpen
  \bibfield  {author} {\bibinfo {author} {\bibfnamefont {A.A.}\ \bibnamefont
  {Hadar}}, \bibinfo {author} {\bibfnamefont {D.}~\bibnamefont {Eliraz}},
  \bibinfo {author} {\bibfnamefont {A.}~\bibnamefont {Lazarovits}}, \bibinfo
  {author} {\bibfnamefont {U.}~\bibnamefont {Alyagon}}, \ and\ \bibinfo
  {author} {\bibfnamefont {A.}~\bibnamefont {Zangen}},\ }\bibfield  {title}
  {\enquote {\bibinfo {title} {Using longitudinal exposure to causally link
  smartphone usage to changes in behavior, cognition and right prefrontal
  neural activity},}\ }\href {\doibase 10.1016/j.brs.2015.01.032} {\bibfield
  {journal} {\bibinfo  {journal} {Brain Stimulation}\ }\textbf {\bibinfo
  {volume} {8}},\ \bibinfo {pages} {318} (\bibinfo {year} {2015})}\BibitemShut
  {NoStop}%
\bibitem [{\citenamefont {Loh}\ and\ \citenamefont {Kanai}(2016)}]{loh2015}%
  \BibitemOpen
  \bibfield  {author} {\bibinfo {author} {\bibfnamefont {Kep~Kee}\ \bibnamefont
  {Loh}}\ and\ \bibinfo {author} {\bibfnamefont {Riota}\ \bibnamefont
  {Kanai}},\ }\bibfield  {title} {\enquote {\bibinfo {title} {How has the
  internet reshaped human cognition?}}\ }\href {\doibase
  10.1177/1073858415595005} {\bibfield  {journal} {\bibinfo  {journal} {The
  Neuroscientist}\ }\textbf {\bibinfo {volume} {22}},\ \bibinfo {pages}
  {506--520} (\bibinfo {year} {2016})}\BibitemShut {NoStop}%
\bibitem [{\citenamefont {Wilmer}\ \emph {et~al.}(2017)\citenamefont {Wilmer},
  \citenamefont {Sherman},\ and\ \citenamefont {Chein}}]{wilmer2017}%
  \BibitemOpen
  \bibfield  {author} {\bibinfo {author} {\bibfnamefont {H.H.}\ \bibnamefont
  {Wilmer}}, \bibinfo {author} {\bibfnamefont {L.E.}\ \bibnamefont {Sherman}},
  \ and\ \bibinfo {author} {\bibfnamefont {J.M.}\ \bibnamefont {Chein}},\
  }\bibfield  {title} {\enquote {\bibinfo {title} {Smartphones and cognition: A
  review of research exploring the links between mobile technology habits and
  cognitive functioning},}\ }\href {\doibase 10.3389/fpsyg.2017.00605}
  {\bibfield  {journal} {\bibinfo  {journal} {Frontiers in Psychology}\
  }\textbf {\bibinfo {volume} {8}} (\bibinfo {year} {2017}),\
  10.3389/fpsyg.2017.00605}\BibitemShut {NoStop}%
\bibitem [{\citenamefont {Falaki}\ \emph {et~al.}(2010)\citenamefont {Falaki},
  \citenamefont {Mahajan}, \citenamefont {Kandula}, \citenamefont
  {Lymberopoulos}, \citenamefont {Govindan},\ and\ \citenamefont
  {Estrin}}]{falaki2010diversity}%
  \BibitemOpen
  \bibfield  {author} {\bibinfo {author} {\bibfnamefont {Hossein}\ \bibnamefont
  {Falaki}}, \bibinfo {author} {\bibfnamefont {Ratul}\ \bibnamefont {Mahajan}},
  \bibinfo {author} {\bibfnamefont {Srikanth}\ \bibnamefont {Kandula}},
  \bibinfo {author} {\bibfnamefont {Dimitrios}\ \bibnamefont {Lymberopoulos}},
  \bibinfo {author} {\bibfnamefont {Ramesh}\ \bibnamefont {Govindan}}, \ and\
  \bibinfo {author} {\bibfnamefont {Deborah}\ \bibnamefont {Estrin}},\
  }\bibfield  {title} {\enquote {\bibinfo {title} {Diversity in smartphone
  usage},}\ }in\ \href {\doibase 10.1145/1814433.1814453} {\emph {\bibinfo
  {booktitle} {Proceedings of the 8th International Conference on Mobile
  Systems, Applications, and Services}}},\ \bibinfo {series and number}
  {MobiSys '10}\ (\bibinfo  {publisher} {ACM},\ \bibinfo {address} {New York,
  NY, USA},\ \bibinfo {year} {2010})\ pp.\ \bibinfo {pages}
  {179--194}\BibitemShut {NoStop}%
\bibitem [{\citenamefont {Marquez}\ \emph {et~al.}(2017)\citenamefont
  {Marquez}, \citenamefont {Gramaglia}, \citenamefont {Fiore}, \citenamefont
  {Banchs}, \citenamefont {Ziemlicki},\ and\ \citenamefont
  {Smoreda}}]{marquez2017not}%
  \BibitemOpen
  \bibfield  {author} {\bibinfo {author} {\bibfnamefont {Cristina}\
  \bibnamefont {Marquez}}, \bibinfo {author} {\bibfnamefont {Marco}\
  \bibnamefont {Gramaglia}}, \bibinfo {author} {\bibfnamefont {Marco}\
  \bibnamefont {Fiore}}, \bibinfo {author} {\bibfnamefont {Albert}\
  \bibnamefont {Banchs}}, \bibinfo {author} {\bibfnamefont {Cezary}\
  \bibnamefont {Ziemlicki}}, \ and\ \bibinfo {author} {\bibfnamefont
  {Zbigniew}\ \bibnamefont {Smoreda}},\ }\bibfield  {title} {\enquote {\bibinfo
  {title} {Not all apps are created equal: Analysis of spatiotemporal
  heterogeneity in nationwide mobile service usage},}\ }in\ \href {\doibase
  10.1145/3143361.3143369} {\emph {\bibinfo {booktitle} {Proceedings of the
  13th International Conference on emerging Networking EXperiments and
  Technologies}}}\ (\bibinfo {organization} {ACM},\ \bibinfo {year} {2017})\
  pp.\ \bibinfo {pages} {180--186}\BibitemShut {NoStop}%
\bibitem [{\citenamefont {Sonntag}\ \emph {et~al.}(2013)\citenamefont
  {Sonntag}, \citenamefont {Manner},\ and\ \citenamefont
  {Schulte}}]{sonntag2013netradar}%
  \BibitemOpen
  \bibfield  {author} {\bibinfo {author} {\bibfnamefont {Sebastian}\
  \bibnamefont {Sonntag}}, \bibinfo {author} {\bibfnamefont {Jukka}\
  \bibnamefont {Manner}}, \ and\ \bibinfo {author} {\bibfnamefont {Lennart}\
  \bibnamefont {Schulte}},\ }\bibfield  {title} {\enquote {\bibinfo {title}
  {Netradar-measuring the wireless world},}\ }in\ \href@noop {} {\emph
  {\bibinfo {booktitle} {2013 11th International Symposium and Workshops on
  Modeling and Optimization in Mobile, Ad Hoc and Wireless Networks (WiOpt)}}}\
  (\bibinfo {organization} {IEEE},\ \bibinfo {year} {2013})\ pp.\ \bibinfo
  {pages} {29--34}\BibitemShut {NoStop}%
\bibitem [{\citenamefont {Peltonen}\ \emph {et~al.}(2018)\citenamefont
  {Peltonen}, \citenamefont {Lagerspetz}, \citenamefont {Hamberg},
  \citenamefont {Mehrotra}, \citenamefont {Musolesi}, \citenamefont {Nurmi},\
  and\ \citenamefont {Tarkoma}}]{peltonen2018hidden}%
  \BibitemOpen
  \bibfield  {author} {\bibinfo {author} {\bibfnamefont {Ella}\ \bibnamefont
  {Peltonen}}, \bibinfo {author} {\bibfnamefont {Eemil}\ \bibnamefont
  {Lagerspetz}}, \bibinfo {author} {\bibfnamefont {Jonatan}\ \bibnamefont
  {Hamberg}}, \bibinfo {author} {\bibfnamefont {Abhinav}\ \bibnamefont
  {Mehrotra}}, \bibinfo {author} {\bibfnamefont {Mirco}\ \bibnamefont
  {Musolesi}}, \bibinfo {author} {\bibfnamefont {Petteri}\ \bibnamefont
  {Nurmi}}, \ and\ \bibinfo {author} {\bibfnamefont {Sasu}\ \bibnamefont
  {Tarkoma}},\ }\bibfield  {title} {\enquote {\bibinfo {title} {The hidden
  image of mobile apps: Geographic, demographic, and cultural factors in mobile
  usage},}\ }in\ \href {\doibase 10.1145/3229434.3229474} {\emph {\bibinfo
  {booktitle} {Proceedings of the 20th International Conference on
  Human-Computer Interaction with Mobile Devices and Services}}},\ \bibinfo
  {series and number} {MobileHCI '18}\ (\bibinfo  {publisher} {ACM},\ \bibinfo
  {address} {New York, NY, USA},\ \bibinfo {year} {2018})\BibitemShut {NoStop}%
\bibitem [{\citenamefont {Shin}\ \emph {et~al.}(2012)\citenamefont {Shin},
  \citenamefont {Hong},\ and\ \citenamefont {Dey}}]{shin2012understanding}%
  \BibitemOpen
  \bibfield  {author} {\bibinfo {author} {\bibfnamefont {Choonsung}\
  \bibnamefont {Shin}}, \bibinfo {author} {\bibfnamefont {Jin-Hyuk}\
  \bibnamefont {Hong}}, \ and\ \bibinfo {author} {\bibfnamefont {Anind~K.}\
  \bibnamefont {Dey}},\ }\bibfield  {title} {\enquote {\bibinfo {title}
  {Understanding and prediction of mobile application usage for smart
  phones},}\ }in\ \href {\doibase 10.1145/2370216.2370243} {\emph {\bibinfo
  {booktitle} {Proceedings of the 2012 ACM Conference on Ubiquitous
  Computing}}},\ \bibinfo {series and number} {UbiComp '12}\ (\bibinfo
  {publisher} {ACM},\ \bibinfo {address} {New York, NY, USA},\ \bibinfo {year}
  {2012})\ pp.\ \bibinfo {pages} {173--182}\BibitemShut {NoStop}%
\bibitem [{\citenamefont {{Yang}}\ \emph {et~al.}(2016)\citenamefont {{Yang}},
  \citenamefont {{Yuan}}, \citenamefont {{Wang}}, \citenamefont {{Zhang}},\
  and\ \citenamefont {{Zeng}}}]{yang2016apps}%
  \BibitemOpen
  \bibfield  {author} {\bibinfo {author} {\bibfnamefont {L.}~\bibnamefont
  {{Yang}}}, \bibinfo {author} {\bibfnamefont {M.}~\bibnamefont {{Yuan}}},
  \bibinfo {author} {\bibfnamefont {W.}~\bibnamefont {{Wang}}}, \bibinfo
  {author} {\bibfnamefont {Q.}~\bibnamefont {{Zhang}}}, \ and\ \bibinfo
  {author} {\bibfnamefont {J.}~\bibnamefont {{Zeng}}},\ }\bibfield  {title}
  {\enquote {\bibinfo {title} {Apps on the move: A fine-grained analysis of
  usage behavior of mobile apps},}\ }in\ \href {\doibase
  10.1109/INFOCOM.2016.7524464} {\emph {\bibinfo {booktitle} {IEEE INFOCOM 2016
  - The 35th Annual IEEE International Conference on Computer
  Communications}}}\ (\bibinfo {year} {2016})\ pp.\ \bibinfo {pages}
  {1--9}\BibitemShut {NoStop}%
\bibitem [{\citenamefont {Yu}\ \emph {et~al.}(2018)\citenamefont {Yu},
  \citenamefont {Li}, \citenamefont {Xu}, \citenamefont {Zhang},\ and\
  \citenamefont {Kostakos}}]{yu2018smartphone}%
  \BibitemOpen
  \bibfield  {author} {\bibinfo {author} {\bibfnamefont {Donghan}\ \bibnamefont
  {Yu}}, \bibinfo {author} {\bibfnamefont {Yong}\ \bibnamefont {Li}}, \bibinfo
  {author} {\bibfnamefont {Fengli}\ \bibnamefont {Xu}}, \bibinfo {author}
  {\bibfnamefont {Pengyu}\ \bibnamefont {Zhang}}, \ and\ \bibinfo {author}
  {\bibfnamefont {Vassilis}\ \bibnamefont {Kostakos}},\ }\bibfield  {title}
  {\enquote {\bibinfo {title} {Smartphone app usage prediction using points of
  interest},}\ }\href@noop {} {\bibfield  {journal} {\bibinfo  {journal}
  {Proceedings of the ACM on Interactive, Mobile, Wearable and Ubiquitous
  Technologies}\ }\textbf {\bibinfo {volume} {1}},\ \bibinfo {pages} {174}
  (\bibinfo {year} {2018})}\BibitemShut {NoStop}%
\bibitem [{\citenamefont {Karatzoglou}\ \emph {et~al.}(2012)\citenamefont
  {Karatzoglou}, \citenamefont {Baltrunas}, \citenamefont {Church},\ and\
  \citenamefont {B{\"o}hmer}}]{karatzoglou2012climbing}%
  \BibitemOpen
  \bibfield  {author} {\bibinfo {author} {\bibfnamefont {Alexandros}\
  \bibnamefont {Karatzoglou}}, \bibinfo {author} {\bibfnamefont {Linas}\
  \bibnamefont {Baltrunas}}, \bibinfo {author} {\bibfnamefont {Karen}\
  \bibnamefont {Church}}, \ and\ \bibinfo {author} {\bibfnamefont {Matthias}\
  \bibnamefont {B{\"o}hmer}},\ }\bibfield  {title} {\enquote {\bibinfo {title}
  {Climbing the app wall: enabling mobile app discovery through context-aware
  recommendations},}\ }in\ \href {\doibase 10.1145/2396761.2398683} {\emph
  {\bibinfo {booktitle} {Proceedings of the 21st ACM international conference
  on Information and knowledge management}}}\ (\bibinfo {organization} {ACM},\
  \bibinfo {year} {2012})\ pp.\ \bibinfo {pages} {2527--2530}\BibitemShut
  {NoStop}%
\bibitem [{\citenamefont {Do}\ and\ \citenamefont
  {Gatica-Perez}(2010)}]{do2010their}%
  \BibitemOpen
  \bibfield  {author} {\bibinfo {author} {\bibfnamefont {Trinh-Minh-Tri}\
  \bibnamefont {Do}}\ and\ \bibinfo {author} {\bibfnamefont {Daniel}\
  \bibnamefont {Gatica-Perez}},\ }\bibfield  {title} {\enquote {\bibinfo
  {title} {By their apps you shall understand them: Mining large-scale patterns
  of mobile phone usage},}\ }in\ \href {\doibase 10.1145/1899475.1899502}
  {\emph {\bibinfo {booktitle} {Proceedings of the 9th International Conference
  on Mobile and Ubiquitous Multimedia}}},\ \bibinfo {series and number} {MUM
  '10}\ (\bibinfo  {publisher} {ACM},\ \bibinfo {address} {New York, NY, USA},\
  \bibinfo {year} {2010})\ pp.\ \bibinfo {pages} {27:1--27:10}\BibitemShut
  {NoStop}%
\bibitem [{\citenamefont {Zhao}\ \emph {et~al.}(2016)\citenamefont {Zhao},
  \citenamefont {Ramos}, \citenamefont {Tao}, \citenamefont {Jiang},
  \citenamefont {Li}, \citenamefont {Wu}, \citenamefont {Pan},\ and\
  \citenamefont {Dey}}]{zhao2016discovering}%
  \BibitemOpen
  \bibfield  {author} {\bibinfo {author} {\bibfnamefont {Sha}\ \bibnamefont
  {Zhao}}, \bibinfo {author} {\bibfnamefont {Julian}\ \bibnamefont {Ramos}},
  \bibinfo {author} {\bibfnamefont {Jianrong}\ \bibnamefont {Tao}}, \bibinfo
  {author} {\bibfnamefont {Ziwen}\ \bibnamefont {Jiang}}, \bibinfo {author}
  {\bibfnamefont {Shijian}\ \bibnamefont {Li}}, \bibinfo {author}
  {\bibfnamefont {Zhaohui}\ \bibnamefont {Wu}}, \bibinfo {author}
  {\bibfnamefont {Gang}\ \bibnamefont {Pan}}, \ and\ \bibinfo {author}
  {\bibfnamefont {Anind~K.}\ \bibnamefont {Dey}},\ }\bibfield  {title}
  {\enquote {\bibinfo {title} {Discovering different kinds of smartphone users
  through their application usage behaviors},}\ }in\ \href {\doibase
  10.1145/2971648.2971696} {\emph {\bibinfo {booktitle} {Proceedings of the
  2016 ACM International Joint Conference on Pervasive and Ubiquitous
  Computing}}},\ \bibinfo {series and number} {UbiComp '16}\ (\bibinfo
  {publisher} {ACM},\ \bibinfo {address} {New York, NY, USA},\ \bibinfo {year}
  {2016})\ pp.\ \bibinfo {pages} {498--509}\BibitemShut {NoStop}%
\bibitem [{\citenamefont {Aggarwal}\ \emph {et~al.}(2014)\citenamefont
  {Aggarwal}, \citenamefont {Halepovic}, \citenamefont {Pang}, \citenamefont
  {Venkataraman},\ and\ \citenamefont {Yan}}]{aggarwal2014prometheus}%
  \BibitemOpen
  \bibfield  {author} {\bibinfo {author} {\bibfnamefont {Vaneet}\ \bibnamefont
  {Aggarwal}}, \bibinfo {author} {\bibfnamefont {Emir}\ \bibnamefont
  {Halepovic}}, \bibinfo {author} {\bibfnamefont {Jeffrey}\ \bibnamefont
  {Pang}}, \bibinfo {author} {\bibfnamefont {Shobha}\ \bibnamefont
  {Venkataraman}}, \ and\ \bibinfo {author} {\bibfnamefont {He}~\bibnamefont
  {Yan}},\ }\bibfield  {title} {\enquote {\bibinfo {title} {Prometheus: toward
  quality-of-experience estimation for mobile apps from passive network
  measurements},}\ }in\ \href@noop {} {\emph {\bibinfo {booktitle} {Proceedings
  of the 15th Workshop on Mobile Computing Systems and Applications}}}\
  (\bibinfo {organization} {ACM},\ \bibinfo {year} {2014})\ p.~\bibinfo {pages}
  {18}\BibitemShut {NoStop}%
\bibitem [{\citenamefont {Song}\ \emph
  {et~al.}(2010{\natexlab{a}})\citenamefont {Song}, \citenamefont {Koren},
  \citenamefont {Wang},\ and\ \citenamefont
  {Barab{\'a}si}}]{song2010modelling}%
  \BibitemOpen
  \bibfield  {author} {\bibinfo {author} {\bibfnamefont {Chaoming}\
  \bibnamefont {Song}}, \bibinfo {author} {\bibfnamefont {Tal}\ \bibnamefont
  {Koren}}, \bibinfo {author} {\bibfnamefont {Pu}~\bibnamefont {Wang}}, \ and\
  \bibinfo {author} {\bibfnamefont {Albert-L{\'a}szl{\'o}}\ \bibnamefont
  {Barab{\'a}si}},\ }\bibfield  {title} {\enquote {\bibinfo {title} {Modelling
  the scaling properties of human mobility},}\ }\href {\doibase
  10.1038/nphys1760} {\bibfield  {journal} {\bibinfo  {journal} {Nature
  Physics}\ }\textbf {\bibinfo {volume} {6}},\ \bibinfo {pages} {818} (\bibinfo
  {year} {2010}{\natexlab{a}})}\BibitemShut {NoStop}%
\bibitem [{\citenamefont {Song}\ \emph
  {et~al.}(2010{\natexlab{b}})\citenamefont {Song}, \citenamefont {Qu},
  \citenamefont {Blumm},\ and\ \citenamefont {Barab{\'a}si}}]{song2010limits}%
  \BibitemOpen
  \bibfield  {author} {\bibinfo {author} {\bibfnamefont {Chaoming}\
  \bibnamefont {Song}}, \bibinfo {author} {\bibfnamefont {Zehui}\ \bibnamefont
  {Qu}}, \bibinfo {author} {\bibfnamefont {Nicholas}\ \bibnamefont {Blumm}}, \
  and\ \bibinfo {author} {\bibfnamefont {Albert-L{\'a}szl{\'o}}\ \bibnamefont
  {Barab{\'a}si}},\ }\bibfield  {title} {\enquote {\bibinfo {title} {Limits of
  predictability in human mobility},}\ }\href {\doibase
  10.1126/science.1177170} {\bibfield  {journal} {\bibinfo  {journal}
  {Science}\ }\textbf {\bibinfo {volume} {327}},\ \bibinfo {pages} {1018--1021}
  (\bibinfo {year} {2010}{\natexlab{b}})}\BibitemShut {NoStop}%
\bibitem [{\citenamefont {Katz}\ and\ \citenamefont
  {Shapiro}(1994)}]{katz1994systems}%
  \BibitemOpen
  \bibfield  {author} {\bibinfo {author} {\bibfnamefont {Michael~L}\
  \bibnamefont {Katz}}\ and\ \bibinfo {author} {\bibfnamefont {Carl}\
  \bibnamefont {Shapiro}},\ }\bibfield  {title} {\enquote {\bibinfo {title}
  {Systems competition and network effects},}\ }\href {\doibase
  10.1257/jep.8.2.93} {\bibfield  {journal} {\bibinfo  {journal} {Journal of
  economic perspectives}\ }\textbf {\bibinfo {volume} {8}},\ \bibinfo {pages}
  {93--115} (\bibinfo {year} {1994})}\BibitemShut {NoStop}%
\bibitem [{\citenamefont {Pan}\ \emph {et~al.}(2011)\citenamefont {Pan},
  \citenamefont {Aharony},\ and\ \citenamefont {Pentland}}]{pan2011composite}%
  \BibitemOpen
  \bibfield  {author} {\bibinfo {author} {\bibfnamefont {Wei}\ \bibnamefont
  {Pan}}, \bibinfo {author} {\bibfnamefont {Nadav}\ \bibnamefont {Aharony}}, \
  and\ \bibinfo {author} {\bibfnamefont {Alex}\ \bibnamefont {Pentland}},\
  }\bibfield  {title} {\enquote {\bibinfo {title} {Composite social network for
  predicting mobile apps installation.}}\ }in\ \href@noop {} {\emph {\bibinfo
  {booktitle} {AAAI}}},\ \bibinfo {series and number} {\bibinfo {number}
  {7.4}}\ (\bibinfo {year} {2011})\ p.~\bibinfo {pages} {2}\BibitemShut
  {NoStop}%
\bibitem [{\citenamefont {Massey}(1951)}]{ref1}%
  \BibitemOpen
  \bibfield  {author} {\bibinfo {author} {\bibfnamefont {Frank~J.}\
  \bibnamefont {Massey}},\ }\bibfield  {title} {\enquote {\bibinfo {title} {The
  kolmogorov-smirnov test for goodness of fit},}\ }\href
  {http://www.jstor.org/stable/2280095} {\bibfield  {journal} {\bibinfo
  {journal} {Journal of the American Statistical Association}\ }\textbf
  {\bibinfo {volume} {46}},\ \bibinfo {pages} {68--78} (\bibinfo {year}
  {1951})}\BibitemShut {NoStop}%
\bibitem [{\citenamefont {Riefer}\ \emph {et~al.}(2017)\citenamefont {Riefer},
  \citenamefont {Prior}, \citenamefont {Blair}, \citenamefont {Pavey},\ and\
  \citenamefont {Love}}]{riefer2017coherency}%
  \BibitemOpen
  \bibfield  {author} {\bibinfo {author} {\bibfnamefont {Peter~S}\ \bibnamefont
  {Riefer}}, \bibinfo {author} {\bibfnamefont {Rosie}\ \bibnamefont {Prior}},
  \bibinfo {author} {\bibfnamefont {Nicholas}\ \bibnamefont {Blair}}, \bibinfo
  {author} {\bibfnamefont {Giles}\ \bibnamefont {Pavey}}, \ and\ \bibinfo
  {author} {\bibfnamefont {Bradley~C}\ \bibnamefont {Love}},\ }\bibfield
  {title} {\enquote {\bibinfo {title} {Coherency-maximizing exploration in the
  supermarket},}\ }\href@noop {} {\bibfield  {journal} {\bibinfo  {journal}
  {Nature human behaviour}\ }\textbf {\bibinfo {volume} {1}},\ \bibinfo {pages}
  {0017} (\bibinfo {year} {2017})}\BibitemShut {NoStop}%
\bibitem [{\citenamefont {Pappalardo}\ \emph {et~al.}(2015)\citenamefont
  {Pappalardo}, \citenamefont {Simini}, \citenamefont {Rinzivillo},
  \citenamefont {Pedreschi}, \citenamefont {Giannotti},\ and\ \citenamefont
  {Barab{\'a}si}}]{pappalardo2015returners}%
  \BibitemOpen
  \bibfield  {author} {\bibinfo {author} {\bibfnamefont {Luca}\ \bibnamefont
  {Pappalardo}}, \bibinfo {author} {\bibfnamefont {Filippo}\ \bibnamefont
  {Simini}}, \bibinfo {author} {\bibfnamefont {Salvatore}\ \bibnamefont
  {Rinzivillo}}, \bibinfo {author} {\bibfnamefont {Dino}\ \bibnamefont
  {Pedreschi}}, \bibinfo {author} {\bibfnamefont {Fosca}\ \bibnamefont
  {Giannotti}}, \ and\ \bibinfo {author} {\bibfnamefont
  {Albert-L{\'a}szl{\'o}}\ \bibnamefont {Barab{\'a}si}},\ }\bibfield  {title}
  {\enquote {\bibinfo {title} {Returners and explorers dichotomy in human
  mobility},}\ }\href {\doibase 10.1038/ncomms9166} {\bibfield  {journal}
  {\bibinfo  {journal} {Nature communications}\ }\textbf {\bibinfo {volume}
  {6}},\ \bibinfo {pages} {8166} (\bibinfo {year} {2015})}\BibitemShut
  {NoStop}%
\bibitem [{\citenamefont {Toole}\ \emph {et~al.}(2015)\citenamefont {Toole},
  \citenamefont {Herrera-Yaqüe}, \citenamefont {Schneider},\ and\
  \citenamefont {González}}]{doi:10.1098/rsif.2014.1128}%
  \BibitemOpen
  \bibfield  {author} {\bibinfo {author} {\bibfnamefont {Jameson~L.}\
  \bibnamefont {Toole}}, \bibinfo {author} {\bibfnamefont {Carlos}\
  \bibnamefont {Herrera-Yaqüe}}, \bibinfo {author} {\bibfnamefont
  {Christian~M.}\ \bibnamefont {Schneider}}, \ and\ \bibinfo {author}
  {\bibfnamefont {Marta~C.}\ \bibnamefont {González}},\ }\bibfield  {title}
  {\enquote {\bibinfo {title} {Coupling human mobility and social ties},}\
  }\href {\doibase 10.1098/rsif.2014.1128} {\bibfield  {journal} {\bibinfo
  {journal} {Journal of The Royal Society Interface}\ }\textbf {\bibinfo
  {volume} {12}},\ \bibinfo {pages} {20141128} (\bibinfo {year}
  {2015})}\BibitemShut {NoStop}%
\bibitem [{\citenamefont {Rauch}(2018)}]{rauch_2018}%
  \BibitemOpen
  \bibfield  {author} {\bibinfo {author} {\bibfnamefont {Jonathan}\
  \bibnamefont {Rauch}},\ }\href@noop {} {\emph {\bibinfo {title} {The
  happiness curve: why life turns around in middle age}}}\ (\bibinfo
  {publisher} {Green Tree, Bloomsbury Publishing, Plc},\ \bibinfo {address}
  {New York, NY},\ \bibinfo {year} {2018})\BibitemShut {NoStop}%
\bibitem [{\citenamefont {Hill}\ and\ \citenamefont {Dunbar}(2003)}]{Hill2003}%
  \BibitemOpen
  \bibfield  {author} {\bibinfo {author} {\bibfnamefont {Russell~A.}\
  \bibnamefont {Hill}}\ and\ \bibinfo {author} {\bibfnamefont {Robin~I.M.}\
  \bibnamefont {Dunbar}},\ }\bibfield  {title} {\enquote {\bibinfo {title}
  {Social network size in humans},}\ }\href {\doibase
  10.1007/s12110-003-1016-y} {\bibfield  {journal} {\bibinfo  {journal} {Human
  Nature}\ }\textbf {\bibinfo {volume} {14}},\ \bibinfo {pages} {53--72}
  (\bibinfo {year} {2003})}\BibitemShut {NoStop}%
\bibitem [{\citenamefont {Wrzus}\ \emph {et~al.}(2013)\citenamefont {Wrzus},
  \citenamefont {H{\"a}nel}, \citenamefont {Wagner},\ and\ \citenamefont
  {Neyer}}]{wrzus2013social}%
  \BibitemOpen
  \bibfield  {author} {\bibinfo {author} {\bibfnamefont {Cornelia}\
  \bibnamefont {Wrzus}}, \bibinfo {author} {\bibfnamefont {Martha}\
  \bibnamefont {H{\"a}nel}}, \bibinfo {author} {\bibfnamefont {Jenny}\
  \bibnamefont {Wagner}}, \ and\ \bibinfo {author} {\bibfnamefont {Franz~J}\
  \bibnamefont {Neyer}},\ }\bibfield  {title} {\enquote {\bibinfo {title}
  {Social network changes and life events across the life span: a
  meta-analysis.}}\ }\href {\doibase 10.1037/a0028601} {\bibfield  {journal}
  {\bibinfo  {journal} {Psychological bulletin}\ }\textbf {\bibinfo {volume}
  {139}},\ \bibinfo {pages} {53} (\bibinfo {year} {2013})}\BibitemShut
  {NoStop}%
\bibitem [{\citenamefont {Holtzman}\ \emph {et~al.}(2004)\citenamefont
  {Holtzman}, \citenamefont {Rebok}, \citenamefont {Saczynski}, \citenamefont
  {Kouzis}, \citenamefont {Wilcox~Doyle},\ and\ \citenamefont
  {Eaton}}]{holtzman2004social}%
  \BibitemOpen
  \bibfield  {author} {\bibinfo {author} {\bibfnamefont {Ronald~E}\
  \bibnamefont {Holtzman}}, \bibinfo {author} {\bibfnamefont {George~W}\
  \bibnamefont {Rebok}}, \bibinfo {author} {\bibfnamefont {Jane~S}\
  \bibnamefont {Saczynski}}, \bibinfo {author} {\bibfnamefont {Anthony~C}\
  \bibnamefont {Kouzis}}, \bibinfo {author} {\bibfnamefont {Kathryn}\
  \bibnamefont {Wilcox~Doyle}}, \ and\ \bibinfo {author} {\bibfnamefont
  {William~W}\ \bibnamefont {Eaton}},\ }\bibfield  {title} {\enquote {\bibinfo
  {title} {Social network characteristics and cognition in middle-aged and
  older adults},}\ }\href@noop {} {\bibfield  {journal} {\bibinfo  {journal}
  {The Journals of Gerontology Series B: Psychological Sciences and Social
  Sciences}\ }\textbf {\bibinfo {volume} {59}},\ \bibinfo {pages} {P278--P284}
  (\bibinfo {year} {2004})}\BibitemShut {NoStop}%
\bibitem [{\citenamefont {Saram{\"a}ki}\ \emph {et~al.}(2014)\citenamefont
  {Saram{\"a}ki}, \citenamefont {Leicht}, \citenamefont {L{\'o}pez},
  \citenamefont {Roberts}, \citenamefont {Reed-Tsochas},\ and\ \citenamefont
  {Dunbar}}]{Saramaki942}%
  \BibitemOpen
  \bibfield  {author} {\bibinfo {author} {\bibfnamefont {Jari}\ \bibnamefont
  {Saram{\"a}ki}}, \bibinfo {author} {\bibfnamefont {E.~A.}\ \bibnamefont
  {Leicht}}, \bibinfo {author} {\bibfnamefont {Eduardo}\ \bibnamefont
  {L{\'o}pez}}, \bibinfo {author} {\bibfnamefont {Sam G.~B.}\ \bibnamefont
  {Roberts}}, \bibinfo {author} {\bibfnamefont {Felix}\ \bibnamefont
  {Reed-Tsochas}}, \ and\ \bibinfo {author} {\bibfnamefont {Robin I.~M.}\
  \bibnamefont {Dunbar}},\ }\bibfield  {title} {\enquote {\bibinfo {title}
  {Persistence of social signatures in human communication},}\ }\href {\doibase
  10.1073/pnas.1308540110} {\bibfield  {journal} {\bibinfo  {journal}
  {Proceedings of the National Academy of Sciences}\ }\textbf {\bibinfo
  {volume} {111}},\ \bibinfo {pages} {942--947} (\bibinfo {year} {2014})},\
  \Eprint {http://arxiv.org/abs/http://www.pnas.org/content/111/3/942.full.pdf}
  {http://www.pnas.org/content/111/3/942.full.pdf} \BibitemShut {NoStop}%
\bibitem [{\citenamefont {Hariharan}\ and\ \citenamefont
  {Toyama}(2004)}]{hariharan2004project}%
  \BibitemOpen
  \bibfield  {author} {\bibinfo {author} {\bibfnamefont {Ramaswamy}\
  \bibnamefont {Hariharan}}\ and\ \bibinfo {author} {\bibfnamefont {Kentaro}\
  \bibnamefont {Toyama}},\ }\bibfield  {title} {\enquote {\bibinfo {title}
  {Project lachesis: Parsing and modeling location histories},}\ }in\
  \href@noop {} {\emph {\bibinfo {booktitle} {Geographic Information
  Science}}},\ \bibinfo {editor} {edited by\ \bibinfo {editor} {\bibfnamefont
  {Max~J.}\ \bibnamefont {Egenhofer}}, \bibinfo {editor} {\bibfnamefont
  {Christian}\ \bibnamefont {Freksa}}, \ and\ \bibinfo {editor} {\bibfnamefont
  {Harvey~J.}\ \bibnamefont {Miller}}}\ (\bibinfo  {publisher} {Springer Berlin
  Heidelberg},\ \bibinfo {address} {Berlin, Heidelberg},\ \bibinfo {year}
  {2004})\ pp.\ \bibinfo {pages} {106--124}\BibitemShut {NoStop}%
\bibitem [{\citenamefont {Ester}\ \emph {et~al.}(1996)\citenamefont {Ester},
  \citenamefont {Kriegel}, \citenamefont {Sander},\ and\ \citenamefont
  {Xu}}]{ester1996density}%
  \BibitemOpen
  \bibfield  {author} {\bibinfo {author} {\bibfnamefont {Martin}\ \bibnamefont
  {Ester}}, \bibinfo {author} {\bibfnamefont {Hans-Peter}\ \bibnamefont
  {Kriegel}}, \bibinfo {author} {\bibfnamefont {J\"{o}rg}\ \bibnamefont
  {Sander}}, \ and\ \bibinfo {author} {\bibfnamefont {Xiaowei}\ \bibnamefont
  {Xu}},\ }\bibfield  {title} {\enquote {\bibinfo {title} {A density-based
  algorithm for discovering clusters a density-based algorithm for discovering
  clusters in large spatial databases with noise},}\ }in\ \href
  {http://dl.acm.org/citation.cfm?id=3001460.3001507} {\emph {\bibinfo
  {booktitle} {Proceedings of the Second International Conference on Knowledge
  Discovery and Data Mining}}},\ \bibinfo {series and number} {KDD'96}\
  (\bibinfo  {publisher} {AAAI Press},\ \bibinfo {year} {1996})\ pp.\ \bibinfo
  {pages} {226--231}\BibitemShut {NoStop}%
\bibitem [{\citenamefont {Kendall}(1938)}]{kendall1938new}%
  \BibitemOpen
  \bibfield  {author} {\bibinfo {author} {\bibfnamefont {Maurice~G}\
  \bibnamefont {Kendall}},\ }\bibfield  {title} {\enquote {\bibinfo {title} {A
  new measure of rank correlation},}\ }\href@noop {} {\bibfield  {journal}
  {\bibinfo  {journal} {Biometrika}\ }\textbf {\bibinfo {volume} {30}},\
  \bibinfo {pages} {81--93} (\bibinfo {year} {1938})}\BibitemShut {NoStop}%
\bibitem [{\citenamefont {Kenney}\ and\ \citenamefont
  {Gortmaker}(2017)}]{kenney2017united}%
  \BibitemOpen
  \bibfield  {author} {\bibinfo {author} {\bibfnamefont {Erica~L}\ \bibnamefont
  {Kenney}}\ and\ \bibinfo {author} {\bibfnamefont {Steven~L}\ \bibnamefont
  {Gortmaker}},\ }\bibfield  {title} {\enquote {\bibinfo {title} {United states
  adolescents' television, computer, videogame, smartphone, and tablet use:
  associations with sugary drinks, sleep, physical activity, and obesity},}\
  }\href@noop {} {\bibfield  {journal} {\bibinfo  {journal} {The Journal of
  pediatrics}\ }\textbf {\bibinfo {volume} {182}},\ \bibinfo {pages} {144--149}
  (\bibinfo {year} {2017})}\BibitemShut {NoStop}%
\bibitem [{\citenamefont {Katevas}\ \emph {et~al.}(2018)\citenamefont
  {Katevas}, \citenamefont {Arapakis},\ and\ \citenamefont
  {Pielot}}]{katevas2018typical}%
  \BibitemOpen
  \bibfield  {author} {\bibinfo {author} {\bibfnamefont {Kleomenis}\
  \bibnamefont {Katevas}}, \bibinfo {author} {\bibfnamefont {Ioannis}\
  \bibnamefont {Arapakis}}, \ and\ \bibinfo {author} {\bibfnamefont {Martin}\
  \bibnamefont {Pielot}},\ }\bibfield  {title} {\enquote {\bibinfo {title}
  {Typical phone use habits: Intense use does not predict negative
  well-being},}\ }in\ \href {\doibase 10.1145/3229434.3229441} {\emph {\bibinfo
  {booktitle} {Proceedings of the 20th International Conference on
  Human-Computer Interaction with Mobile Devices and Services}}},\ \bibinfo
  {series and number} {MobileHCI '18}\ (\bibinfo  {publisher} {ACM},\ \bibinfo
  {address} {New York, NY, USA},\ \bibinfo {year} {2018})\ pp.\ \bibinfo
  {pages} {11:1--11:13}\BibitemShut {NoStop}%
\end{thebibliography}%
\clearpage

\appendix
\section{The stop location algorithm}
As described in the main manuscript, the \emph{Users Location} data stream is composed of (user ID, date, time, latitude, longitude). 

\mbox{ } \\
\textbf{Stop events.}
From a sequence of ordered time events $T= \left [t_0, t_1, \ldots, t_n \right] \mid t_j \leq t_i$, a corresponding set of GPS locations $R=\left [r_0, r_1, \ldots, r_n\right]$, and a geographical distance function $d(i,j)$, we define a \emph{stop event} as a maximal set of locations $S=\left [r_i, r_{i+1}, \ldots, r_j\right] \mid d(r_i, r_j) < \Delta s \land t_j - t_i \geq \Delta t \ \forall r_i, r_j \in S$. Then the set of \emph{stop events} is $\mathcal{S} = \{S_i \mid S_i \text{ is a stop event} \land r_i \in S_i \land r_j \in S_j \land i<j \}$. To form a \emph{stop event} we heuristically choose to group locations in a time-ordered fashion.
In other words, we aim at finding all those places at most $\Delta s$ meters large were people stopped for at least $\Delta t$ minutes. Each \emph{stop event} is composed by at least two locations and the locations can belong only to at most one \emph{stop event}.

To extract \emph{stop events} we base our method on Hariharan and Toyama's work~\cite{hariharan2004project}. The algorithm is depicted in Algorithm~\ref{alg:one} and can be summarised as follows: for each user, we first order his/her GPS locations by time, followed by selecting groups of GPS sequences with the desired properties to form \emph{stop events}. 
The \texttt{Diameter} function computes the greatest distance between points, while \texttt{Medoid} selects the GPS location with the minimum distance to all other points in the set.

\begin{algorithm}[ht]
\SetAlgoNoLine
\KwIn{Time-ordered list of a user's raw GPS positions $R=[r_0, r_1, \ldots, r_n]$, their time $T=[t_0, t_1, \ldots, t_n]$, a spatial threshold $\Delta s$ and a temporal threshold $\Delta t$.}
\KwOut{The set $S$ of a user's stop events.}
$\aleft$ = 0; $S \leftarrow \emptyset$ \;
\While{$\aleft < n$}{
        $\aright \leftarrow$ minimum $j$ such that $t_j \geq t_{\aleft} + \Delta t$\;
        \If{Diameter($R$, $\aleft$, $j$) $> \Delta s$\;
        }{
        $\aleft\leftarrow \aleft + 1$\;
        }\Else{
        $\aright \leftarrow$ maximum $j$ such that $j \leq n$ and Diameter($R$, $\aleft$, $j$) $ < \Delta s$\;
        $S \leftarrow S \cup$ (Medoid($R$, $\aleft$, $\aright$), $t_{\aleft}$, $t_{\aright}$) \;
        $\aleft \leftarrow \aright + 1$\;
        }
      }
\caption{Algorithm for extracting the stop \emph{events} from GPS sequences.}
\label{alg:one}
\end{algorithm}

The complexity of the \emph{stop event} algorithm~\cite{hariharan2004project} is $\mathcal{O}(n^3)$, because of the repeated \texttt{Diameter} function that computes a distance matrix, whose complexity is $\mathcal{O}(n^2)$. Thus, we make two optimisations to this basic algorithm in order to improve its complexity:
\begin{itemize}
    \item Each time we compute \texttt{Diameter}($R$, $\aleft$, $j$) we cache the computed distance matrix so that we can use it again whenever we need to compute \texttt{Diameter}($R$, $\aleft$, $j+1$). This reduces the complexity to $\mathcal{O}(n^2)$.
    \item We reduce the number of points that are most likely not part of a \emph{stop event}. Thus, we filter out $\forall r_i \mid d(r_{i-1}, r_{i}) < 10\text{m} \land \mid d(r_{i}, r_{i+1}) < 10\text{m}$, but also those $\forall r_j \mid d(r_{j-1}, r_{j}) > \Delta s \land \mid d(r_{j}, r_{j+1}) > \Delta s$. Although simple, this heuristics keep the complexity on average around $\mathcal{O}(n)$ and in the worst case $\mathcal{O}(n^2)$.
\end{itemize}

The \texttt{Diameter} algorithm can be further optimised by converting all coordinates to a Cartesian plane, then finding the smallest convex region containing all the points and finally computing the diameter in linear time between the points of the convex hull. However, in this work we choose to have higher accuracy using the original coordinates and defining $d(i,j)$ as the Haversine great-circle distance between $i$ and $j$.
Given the average radius of the Earth $r$ and two points with latitude and longitude $\varphi_1$, $\varphi_2$ and $\lambda_1$, $\lambda_2$ respectively, the Haversine distance $d$ between them is:
\begin{equation*}
    d = 2 r \arcsin\left(\sqrt{\sin^2\left(\frac{\varphi_2 - \varphi_1}{2}\right) + \cos(\varphi_1) \cos(\varphi_2)\sin^2\left(\frac{\lambda_2 - \lambda_1}{2}\right)}\right)
\end{equation*}
The Haversine distance does not require to project points to a plane, and it is more accurate both in short and long distances.

\mbox{ } \\
\textbf{Stop locations.}
For each user, we define \emph{stop locations} as the sequences of \emph{stop events} that can be considered part of the same place. 
For example: if user A goes many times at the Colosseum in Rome, she could have many \emph{stop events} (\emph{e.g.}, northern entrance, southern entrance) that can be grouped in a unique \emph{stop location} (\emph{i.e.} the Colosseum).
To determine a \emph{stop location} from \emph{stop events} we use the DB-scan~\cite{ester1996density} algorithm that groups points within $\epsilon = \Delta s -5$ meters of distance to form a cluster with at least $minPoints = 1$ \emph{event}. 
The complexity of DB-scan is $\mathcal{O}(n)$. We horizontally scale the computation through different cloud machines thanks to Apache Spark.

Taking as a reference previous work~\cite{alessandretti2018evidence, pappalardo2015returners, hariharan2004project} we choose $\Delta s = 50$ meters and $\Delta t = 15$ minutes. We qualitatively noticed that with $\Delta s = 30$ (same as the error threshold for our data filtering) the stop locations are more noisy. Similarly, $\Delta t = 10$ minutes may form some spurious stop locations.

We select $\epsilon = \Delta s -5$ meters to avoid the creation of an extremely --and incorrect-- long chain of sequential \emph{stop events}. Thus, $\epsilon = 45$ meters.
However, \emph{stop events} and \emph{stop locations} may be very sensible to the $\Delta s$ and $\Delta t$ parameters. Therefore, we repeated our experiments both with $\Delta s = 60$ and $\Delta t=10$ and we found no significant differences. For this reason, in the next Sections we align our discussion to the existing literature and use $\Delta s = 50$ meters and $\Delta t = 15$ minutes.

\section{From applications to mobility}
We investigated the relationship between mobile app usage behaviour and mobility by correlating the capacity, activity, and strategy between app usage and mobility. 
However, temporally aggregated behaviour might hide choices people make at a smaller time scale. 
Thus, we break down people's behaviour on a daily, weekly and monthly basis and test for any trade-off between the number of stop locations and the time spent on different types of apps.
For each user $i$ we compute the number of visited locations $L_i = [l_{i,1}, l_{i,2}, \ldots, l_{i,n}]$, and the time spent on apps $W_i = [w_{i,1}, w_{i,2}, \ldots, w_{i,n}]$ at the chosen level of temporal aggregation. 
Then, we concatenate all $m$ users' behaviours: $\Phi = [L_0, L_1, \ldots, L_m]$ and $\Gamma = [W_0, W_1, \ldots, W_m]$ and test through the Kendall's $\tau$~\cite{kendall1938new} three different variables:
\begin{itemize}
    \item \textbf{Raw}: \emph{if an individual spends more time using apps, does (s)he visit fewer places?} Defined as: $\tau(\Phi, \Gamma)$.
    \item \textbf{Average behavior}: \emph{if an individual spends more time than what other people on average use apps, does his/her mobility decrease?}
    Defined as: $\tau(\Phi', \Gamma')$ with $\Phi' = [L'_0, \ldots, L'_m]$, $\Gamma' = [W'_0, \ldots, W'_m]$, $L'_i = L_i - \overline{L_i}$ and $W'_i = W_i - \overline{W}$. 
    \item \textbf{Individual behavior}: \emph{if an individual spends exceptionally more time than his/her average or baseline on mobile apps, does his/her mobility suffer?} Defined as: $\tau(\Phi'', \Gamma'')$ with $\Phi'' = [L''_0, \ldots, L''_m]$, $\Gamma'' = [W''_0, \ldots, W''_m]$, $L''_i = \frac{L_i - \overline{L_i}}{\omega_{L_i}}$ and $W''_i = \frac{W_i - \overline{W_i}}{\omega_{W_i}}$. 
\end{itemize}
For a set of pairs $(i,j)$ at time $t$, the Kendall rank coefficient measures how much the rank of the pair changed from $t$ to $t+1$.
The coefficient is 1 when the ranks are identical, while it is -1 when they are dissimilar.
In other words, we expect the Kendall's $\tau$ to be positive and high when application usage is very similar to mobility, while we expect it to be negative in the presence of a trade-off between the two domains.

Thus, we compare the app usage and mobility dynamics and look for any trade-off or positive correlation between these two domains. 
A strong negative correlation between the two domains echoes previous studies linking smart-phone addiction to negative outcomes such as obesity~\cite{kenney2017united}, while a strong positive correlation mean people use phones especially when they move, or with a scale-free dynamic.

\Cref{table:tradeoff} summarises the results of such an analysis. As depicted in the Table, we do not find any negative correlation between these variables, which would represent the existence of a trade-off between mobile phone usage and human mobility. On the contrary, we do find a slight positive correlation. In other words, the higher the capacity in mobility, the higher the capacity in the app domain. 

\begin{figure}
    \centering
  \includegraphics[width=\textwidth]{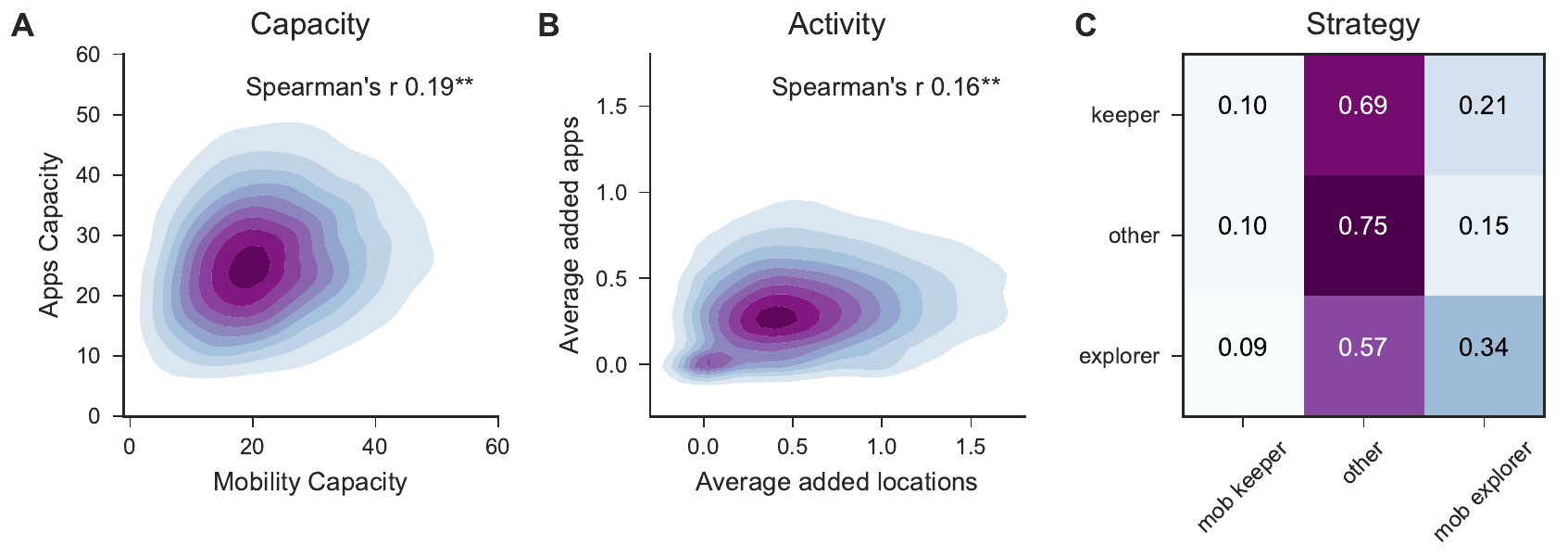}
  \caption{Match between the mobility and digital domain. (A) Density plot showing a positive an significant correlation between users' mobility capacity and application capacity. (B) Density plot between users' mobility and application activity, showing a significant and positive correlation, although grouped in two dense regions. (C) Math between an individual's label in the digital domain and the label assigned in the mobility domain. This confusion matrix shows that labels does not often match across domains. (**) stands for $p$-value $< 0.001$.}
  \label{fig:crosscorr}
\end{figure}

In summary, we find that capacity is positively correlated between the two domains, but users might adopt different strategies in each domain. Empirical results have shown that intense use of the phone does not necessarily predict well-being~\cite{katevas2018typical}. Similarly, our results suggest that people, on average, do not decrease (increase) their physical mobility (as measured by the number of visited places) because of the high (low) phone usage. While the correlation of capacity might be a consequence of the intense phone usage during commuting~\cite{yang2016apps}, we find exciting the fact that there is a difference in the strategies in these two domains for the same user. One could speculate that these two domains reflect different aspects of human behaviour. We leave the investigation of this hypothesis to future work. 

\begin{table}[ht!]
    \ra{1.2}
    \caption{Kendall's $\tau$ correlation between daily, weekly, and monthly number of locations and time spent on applications. The Raw, Average, and Individual's average behaviour are sightly but significantly positive correlated. (**) stands for $p$-value $< 0.001$.}
    \centering
    \begin{tabular}{@{}lrrr@{}}
        \toprule
        \textbf{Granularity} & \textbf{Raw} & \textbf{Average} & \textbf{Individual's average}\\
        \midrule
        Daily   & $0.003^{**}$ & $0.016^{**}$ & $0.043^{**}$ \\
        Weekly  & $0.042^{**}$ & $0.068^{**}$ & $0.070^{**}$ \\
        Monthly & $0.053^{**}$ & $0.076^{**}$ & $0.073^{**}$ \\
         \bottomrule
    \end{tabular}
    \label{table:tradeoff}
\end{table}

\clearpage
\section{Additional figures}
\begin{figure}[ht]
  \includegraphics[width=0.95\textwidth]{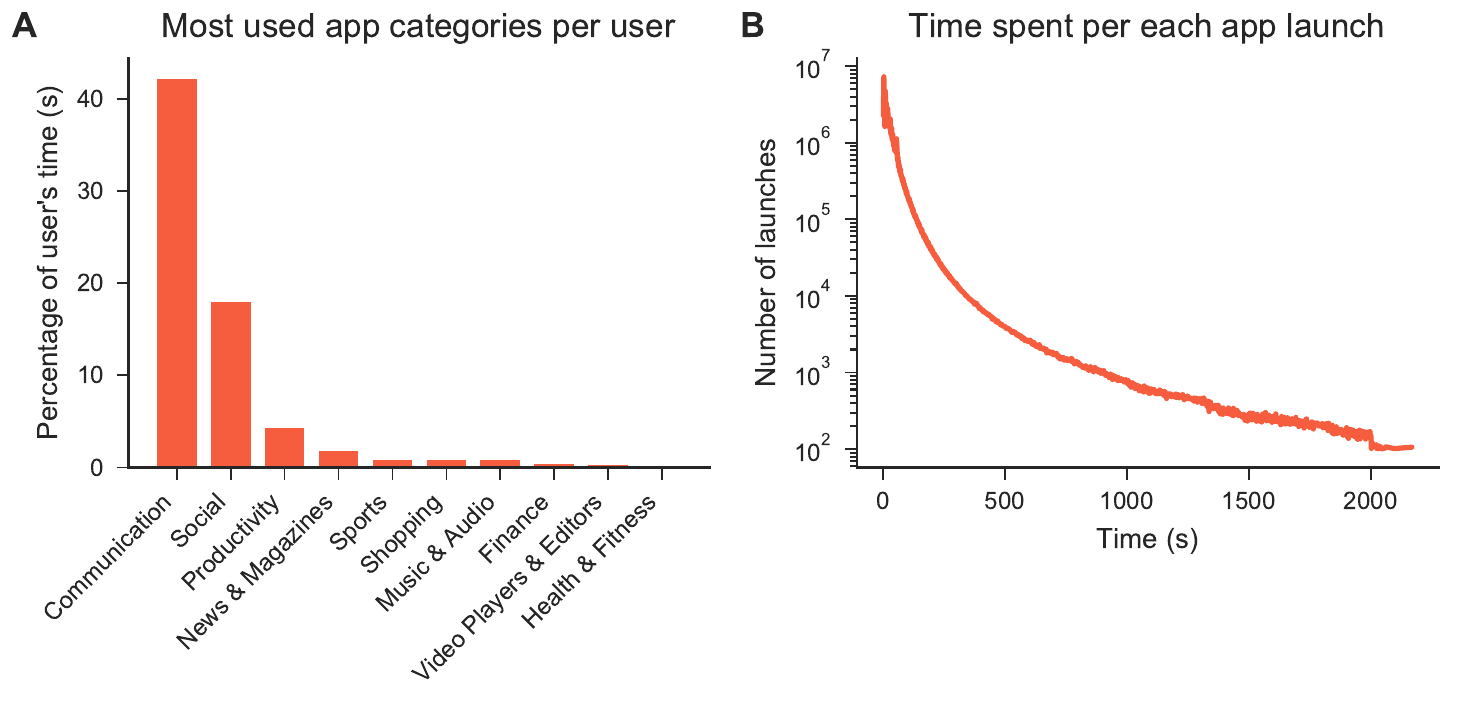}
  \caption{Description of applications launch in our data. (A) The total time spent for each applications' category is very focused on a few categories. (B) PDF Distribution of the time spent on apps per each apps launch.}
  \label{fig:appendix1}
\end{figure}

\begin{figure}[ht]
  \includegraphics[width=0.95\textwidth]{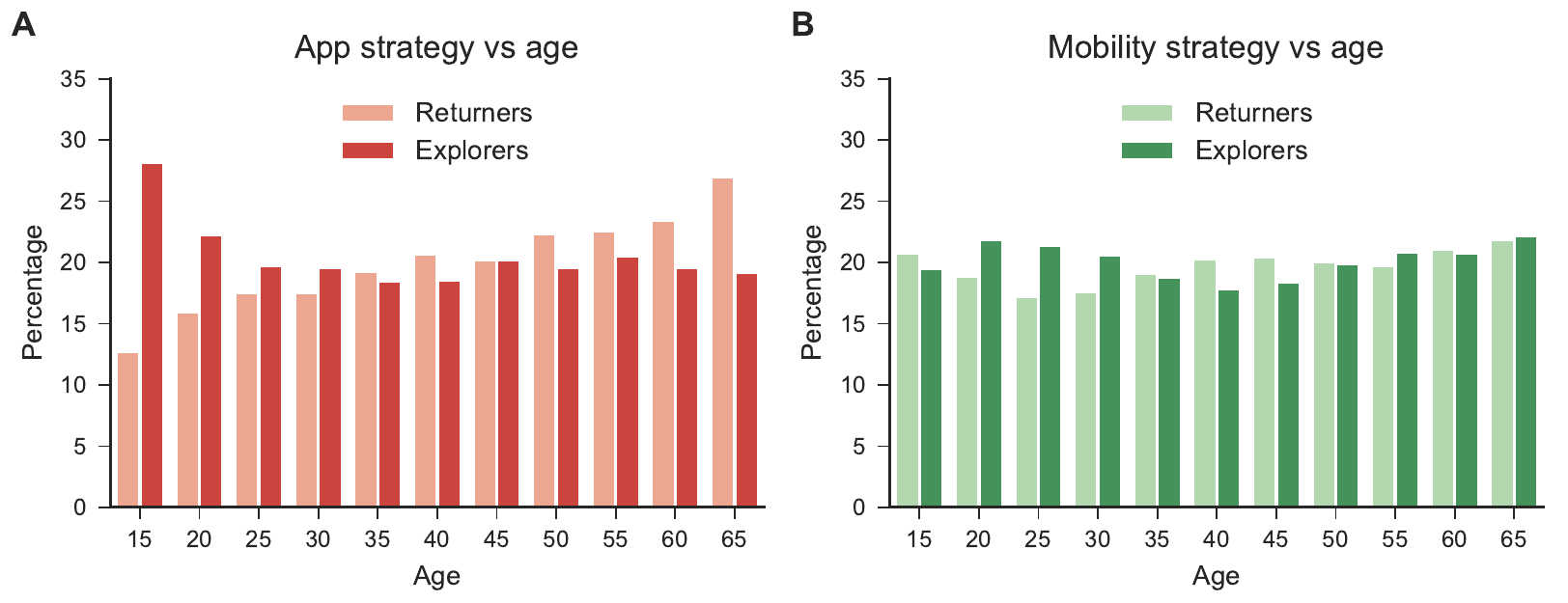}
  \caption{Strategy of people. (A) Percentage of mobility explorers and keepers per each age bin of 5 years. (B) Percentage of app explorers and keepers per each age bin of 5 years.}
  \label{fig:appendix2}
\end{figure}

\begin{figure}[ht]
  \includegraphics[width=0.95\textwidth]{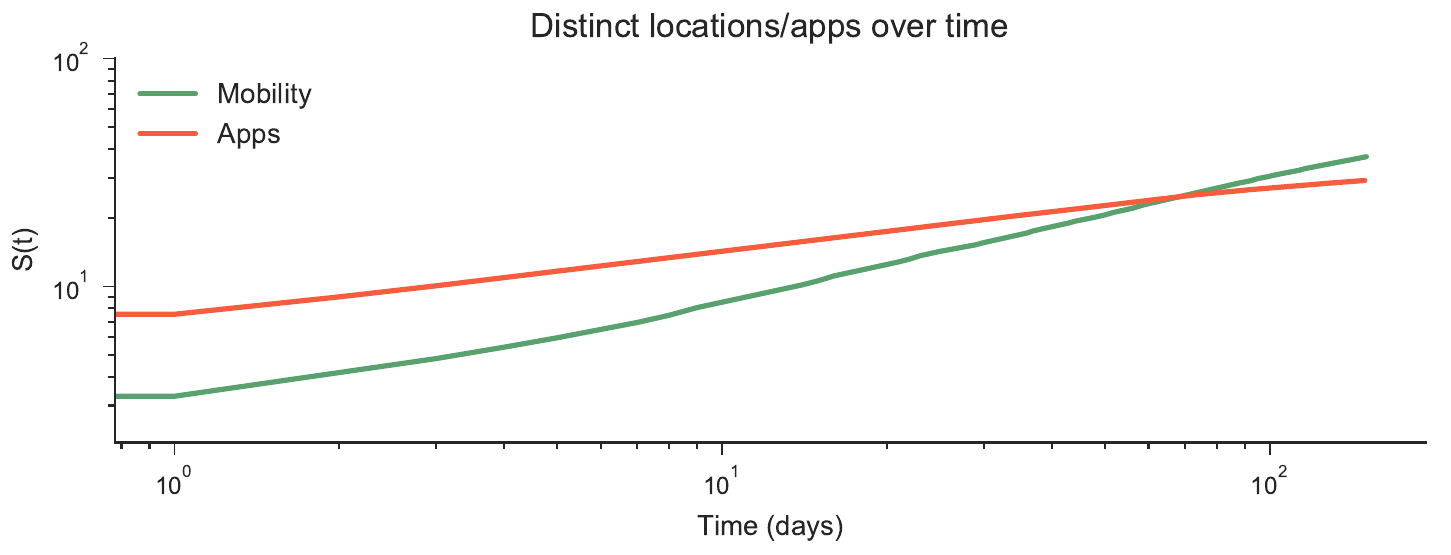}
  \caption{The cumulative number of locations and apps per user over time.}
  \label{fig:appendix_cumsum}
\end{figure}

\begin{figure}[ht]
  \includegraphics[width=\textwidth]{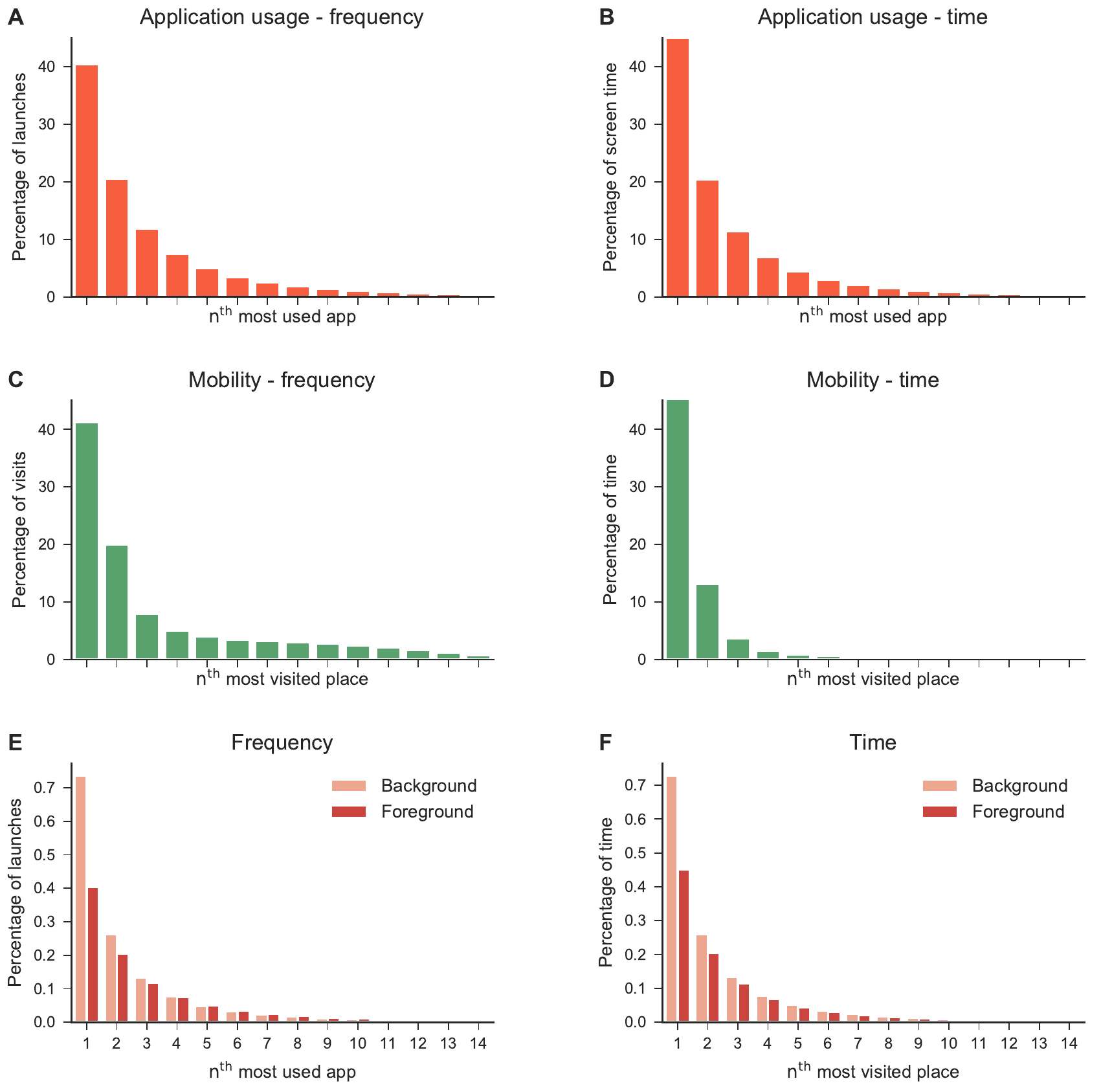}
  \caption{Descriptive statistics in both frequency and time domains, for mobility and application usage. (A) The application usage in terms of frequency. The most used app occupies almost 40\% of users' time. (B) The application usage in terms of frequency. The most used app occupies more than 40\% of users' time. (C) The frequency of visits for the first 15 locations. The most visited place occupies almost 40\% of users' time. (D) The time spent on places for the first 15 locations. (E-F) We compare the frequency of background and foreground apps. The background applications are more skewed than foreground ones in both time and frequency domain.}
  \label{fig:appendix_descriptive}
\end{figure}

\begin{figure}[ht]
  \includegraphics[width=\textwidth]{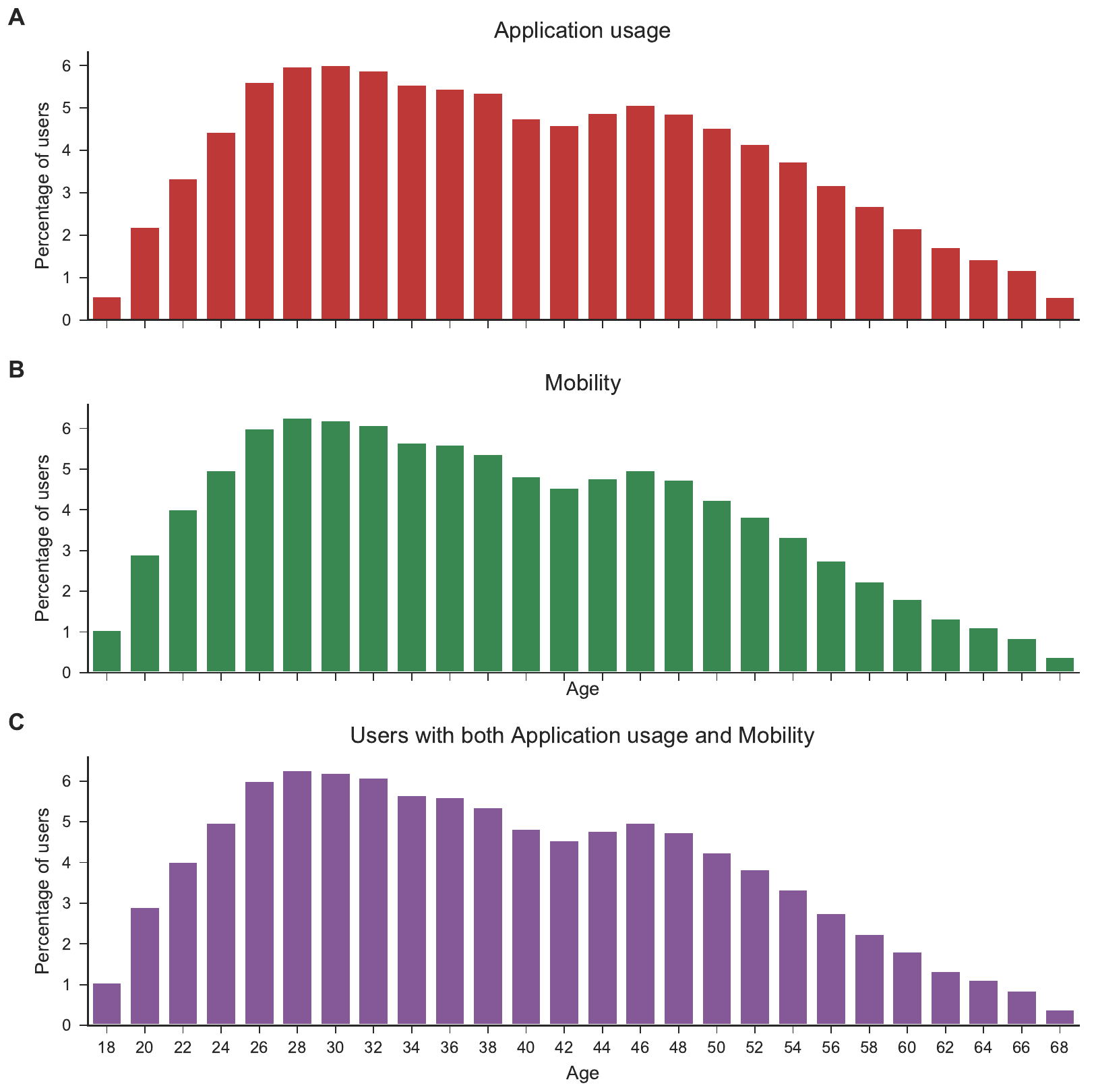}
  \caption{Age distribution in our dataset (in bins of two yers). (A) Distribution of ages for those users where we have mobility information. (B) Distribution of ages for those users where we have application usage information. (c) Distribution of ages for those users where we have both mobility and application usage information.}
  \label{fig:appendix_time}
\end{figure}

\begin{figure}[ht]
  \includegraphics[width=\textwidth]{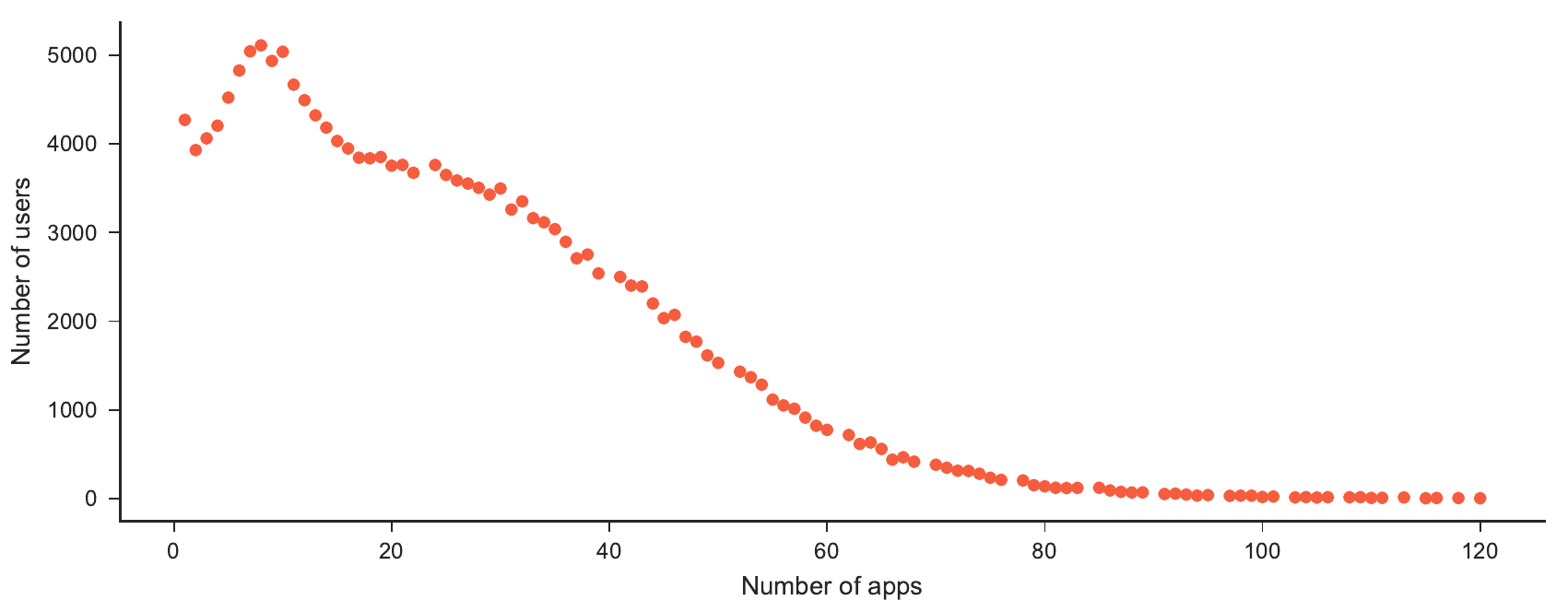}
  \caption{Distribution of the distinct number of apps installed by the users in our data. Note that this number is computed through the applications used at least once in our time period.}
  \label{fig:appendix_number apps}
\end{figure}

\clearpage
\section{Additional tables}
\begin{table}[ht!]
    \ra{1.2}
    \caption{Mobile applications kept for longer time.}
    \centering
    \begin{tabular}{@{}llr@{}}
        \toprule
        \textbf{Package name} & \textbf{Common name} & \textbf{Average n. weeks per user}\\
        \midrule
        \texttt{com.whatsapp}   & Whatsapp & 4.81 \\
        \texttt{com.facebook.katana}   & Facebook & 4.74 \\
        \texttt{com.android.chrome}   & Chrome & 4.52 \\
        \texttt{com.facebook.orca}   & Messanger & 4.41 \\
        \texttt{com.google.android.youtube}   & Youtube & 4.27 \\
        \texttt{com.google.android.apps.maps}   & Maps & 3.82 \\
        \texttt{com.google.android.gm}   & Gmail & 3.18 \\
        \texttt{com.instagram.android}   & Instagram & 3.06 \\
        \texttt{com.sec.android.inputmethod}   & Samsung Keyboard & 2.69 \\
        \texttt{com.sec.android.app.sbrowser}   & Samsung Internet Browser & 2.66 \\
         \bottomrule
    \end{tabular}
    \label{table:mostkept}
\end{table}

\begin{table}[ht!]
    \ra{1.2}
    \caption{Most dropped apps.}
    \centering
    \begin{tabular}{@{}llr@{}}
        \toprule
        \textbf{Package name} & \textbf{Common name} & \textbf{Number drops}\\
        \midrule
        \texttt{com.sec.android.inputmethod}   & Samsung Keyboard &  17928 \\
   \texttt{com.samsung.knox.securefolder}   & Secure Folder &  17310 \\
  \texttt{com.samsung.android.oneconnect}   & SmartThings &  14456 \\
    \texttt{com.google.android.apps.docs}   & Docs &  11895 \\
          \texttt{com.google.android.gm}   & Gmail &   9790 \\
    \texttt{com.google.android.apps.maps}   & Maps &   6526 \\
      \texttt{com.google.android.youtube}   & Youtube &   6325 \\
 \texttt{com.microsoft.office.powerpoint}   & Powerpoint &   4513 \\
       \texttt{com.sec.android.gallery3d}   & Samsung Gallery &   4376 \\
  \texttt{com.google.android.apps.photos}   & Photos &   4259 \\
         \bottomrule
    \end{tabular}
    \label{table:mostdropped}
\end{table}

\begin{table}[ht!]
    \ra{1.2}
    \caption{Categories of apps kept for longer time.}
    \centering
    \begin{tabular}{@{}llr@{}}
        \toprule
        \textbf{Rank} & \textbf{Category name} \\
        \midrule
        1 & Communication \\
        2 & Productivity \\
        3 & Social \\
        4 & Shopping \\
        5 & Travel \\
        6 & Finance \\
        7 & Music \& Audio \\
        8 & Video Players \& Editors \\
        9 & Entertainment \\
        10 & Lifestyle \\
         \bottomrule
    \end{tabular}
    \label{table:catmostkept}
\end{table}

\begin{table}[ht!]
    \ra{1.2}
    \caption{Most dropped categories of apps.}
    \centering
    \begin{tabular}{@{}llr@{}}
        \toprule
        \textbf{Rank} & \textbf{Category name} \\
        \midrule
        1 &    Productivity \\
        2 &       Lifestyle \\
        3 &   Communication \\
        4 &        Shopping \\
        5 &   Entertainment \\
        6 &        Business \\
        7 &  Travel \& Local \\
        8 &          Social \\
        9 &         Finance \\
        10 &   Music \& Audio \\
         \bottomrule
    \end{tabular}
    \label{table:catmostdropped}
\end{table}

\end{document}